\documentclass[preprint2,trackchanges]{aastex62}
\graphicspath{ {./figs/} }
\usepackage{xspace, url, amsmath}

\newcommand\vs{\ensuremath{\mathrm{vs.}}\xspace}
\newcommand{\FSPS}{{\sc FSPS}\xspace}
\newcommand{\pFSPS}{{\tt \textbf{python-fsps}}\xspace}
\newcommand{\CloudyFSPS}{{\tt \textbf{CloudyFSPS}}\xspace}

\newcommand{\Cloudy}{\textsc{Cloudy}\xspace}
\newcommand{\hii}{H\,{\sc ii}\xspace}
\newcommand{\nii}{[N\,{\sc ii}]\xspace}
\newcommand{\sii}{[S\,{\sc ii}]\xspace}
\newcommand{\oiii}{[O\,{\sc iii}]\xspace}

\newcommand{\oi}{[O\,{\sc i}]\xspace}

\newcommand\Lsun{\ensuremath{\,L_{\sun}}\xspace}
\newcommand\Msun{\ensuremath{\,M_{\sun}}\xspace}
\newcommand\Mi{\ensuremath{\,M_{\mathrm{i}}}\xspace}
\newcommand{\ha}{\ensuremath{\mathrm{H\alpha}}\xspace}
\newcommand{\hb}{\ensuremath{\mathrm{H\beta}}\xspace}
\newcommand{\logten}{\ensuremath{\log_{10}}}

\newcommand{\logZeq}[1]{\ensuremath{\logten \mathrm{Z}/\mathrm{Z}_{\sun} = #1}}
\newcommand{\ang}{\ensuremath{\mbox{\AA}}\xspace}

\newcommand{\QH}{\ensuremath{Q_{\mathrm{H}}}\xspace}
\newcommand{\QHe}{\ensuremath{Q_{\mathrm{He}}}\xspace}

\newcommand{\U}{\ensuremath{\mathcal{U}_{0}}\xspace}
\newcommand{\logU}{\ensuremath{\logten \mathcal{U}_0}}
\newcommand{\logUeq}[1]{\ensuremath{\logten \mathcal{U}_0 = #1}}
\newcommand{\kms}{\ensuremath{\;\mathrm{km}\;\mathrm{s}^{-1}}\xspace}
\newcommand{\Myr}{$\,$Myr\xspace}
\newcommand{\Gyr}{$\,$Gyr\xspace}
\newcommand{\logOH}{\ensuremath{\log_{10} (\mathrm{O}/\mathrm{H})}\xspace}
\newcommand{\Teff}{\ensuremath{T_{\mathrm{eff}}}\xspace}
\newcommand{\alphaCN}{\ensuremath{+\alpha\mathrm{CN}}\xspace}

\revised{\today}
\shorttitle{LIER-like emission lines from post-AGB stars}
\shortauthors{Byler et al.}
\begin{document}
\title{Self-consistent predictions for LIER-like emission lines from post-AGB stars}
\author[0000-0002-7392-3637]{Nell Byler}
\affil{Research School of Astronomy and Astrophysics, The Australian National University, ACT, Australia}
\affil{Department of Astronomy, University of Washington, Box 351580, Seattle, WA 98195, USA}

\author[0000-0002-1264-2006]{Julianne J. Dalcanton}
\affil{Department of Astronomy, University of Washington, Box 351580, Seattle, WA 98195, USA}
\author[0000-0002-1590-8551]{Charlie Conroy}
\affil{Department of Astronomy, Harvard University, Cambridge, MA, USA}
\author{Benjamin D. Johnson}
\affil{Department of Astronomy, Harvard University, Cambridge, MA, USA}
\author{Jieun Choi}
\affil{Department of Astronomy, Harvard University, Cambridge, MA, USA}
\author{Aaron Dotter}
\affil{Department of Astronomy, Harvard University, Cambridge, MA, USA}
\author[0000-0001-9306-6049]{Philip Rosenfield}
\affil{Department of Astronomy, Harvard University, Cambridge, MA, USA}
\affil{Eureka Scientific, Inc., 2452 Delmer Street, Oakland CA 94602, USA}
\begin{abstract}

Early type galaxies (ETGs) frequently show emission from warm ionized gas. These Low Ionization Emission Regions (LIERs) were originally attributed to a central, low-luminosity active galactic nuclei. However, the recent discovery of spatially-extended LIER emission suggests ionization by both a central source and an extended component that follows a stellar-like radial distribution. For passively-evolving galaxies with old stellar populations, hot post-Asymptotic Giant Branch (AGB) stars are the only viable extended source of ionizing photons. In this work, we present the first prediction of LIER-like emission from post-AGB stars that is based on fully self-consistent stellar evolution and photoionization models. We show that models where post-AGB stars are the dominant source of ionizing photons reproduce the nebular emission signatures observed in ETGs, including LIER-like emission line ratios in standard optical diagnostic diagrams and \ha equivalent widths of order $0.1-3$\ang. We test the sensitivity of LIER-like emission to the details of post-AGB models, including the mass loss efficiency and convective mixing efficiency, and show that line strengths are relatively insensitive to post-AGB timescale variations. Finally, we examine the UV-optical colors of the models and the stellar populations responsible for the UV-excess observed in some ETGs. We find that allowing as little as 3\% of the HB population to be uniformly distributed to very hot temperatures (30,000\,K) produces realistic UV colors for old, quiescent ETGs.

\end{abstract}
\section{Introduction}\label{sec:intro}

Early type galaxies (ETGs) are usually assumed to be gas-poor, passively-evolving systems. However, we know that ETGs can also harbor reservoirs of diffuse ionized gas \citep[e.g.,][]{Sarzi+2006, Singh+2013, Kehrig+2012, Papaderos+2013}. The source of ionization in these systems has long been debated, with explanations that include the heat transfer from hot to cold gas \citep{Heckman+1981}, shock waves \citep{Koski+1976, Dopita+1995, Allen+2008}, central active galactic nuclei \citep[AGN; ][]{Ferland+1983, Halpern+1983, Heckman+1998, Ho+1999, Kauffmann+2003b, Kewley+2006, Ho+2009}, or evolved stellar sources \citep{Binette+1994, Taniguchi+2000}. Current research supports the view that gas excitation in ETGs on extended, kpc scales is driven by photoionization from evolved stars and fast shocks, with additional contributions from low-luminosity AGN on nuclear scales \citep[e.g., ][]{Pandya+2017}.

The low-ionization emission was originally discovered in the nuclear regions of galaxies and attributed to AGN-related activity \citep[for a recent review, see][]{Filippenko+2003}. However, spatially resolved spectroscopy has revealed that low-ionization emission regions (LIERs) are not always centrally concentrated and instead can be spatially extended over ${\sim}\mathrm{kpc}$ scales \citep{Goudfrooij+1994, Maccehetto+1996, Sharp+2010, Kehrig+2012, Papaderos+2013, Singh+2013, James+2015, Belfiore+2016, Gomes+2016}. While central LIER emission is likely still attributable to low-luminosity AGN activity \citep[e.g.,][]{Kormendy+2013}, it has been suggested that hot evolved stars, which are the primary source of ionizing photons when main sequence stars are absent, are responsible for the spatially-extended LIER emission \citep{Binette+1994, Stasinska+2008, Sarzi+2010, Yan+2012, Woods+2014, Johansson+2016}.

There are two main candidates for non-main sequence (MS) stars that are hot enough while also being either luminous or numerous enough to power significant ionization. The first are extreme horizontal branch (EHB) stars, a sub-group of HB stars, the evolutionary stage immediately following the Red Giant Branch (RGB) phase. Stars with initial masses between 0.8 and 2\Msun undergo a helium flash and begin core helium fusion on the HB. However, the subset of stars that experience more mass loss during their RGB evolution begin their HB phase with less massive hydrogen shells, producing hot or ``extreme'' HB stars. These stars are relatively faint when compared to other possible ionizing sources ($50-100$\Lsun) but are very numerous and relatively long-lived, with lifetimes $\sim 10^8$ years.

The second stellar candidate is post-Asymptotic Giant Branch (AGB) stars, which are stars with initial masses between 0.8 and 8\Msun that have left the AGB and are evolving horizontally across the HR diagram towards very hot temperatures ($\sim10^5$ K) before cooling and fading to become white dwarfs. Although they are not as long-lived as EHB stars ($\sim10^3-10^7$ years), post-AGB stars are much more luminous ($L_{\mathrm{bol}} = 10^{2-4}\Lsun$) and possibly hotter (see reviews in \citealp{OConnell+1999, Grevesse+2010}; also \citealp{Bressan+2012}).

Between extreme HB and post-AGB stars, post-AGB stars are the more promising source of ionizing photons. Nominally, blue HB (BHB) stars have temperatures between 12,000 and 20,000\,K, which is not hot enough to ionize hydrogen. Extreme HB (EHB) stars have temperatures in excess of 20,000\,K, and are hot enough to ionize hydrogen. However, HB morphology varies considerably from system to system, and significant EHB populations are not a widespread phenomena \citep[see recent review in][]{Catelan+2009}. While EHB stars may contribute to the ionizing photon budget in ETGs, these stars are not likely to be the dominant source of ionizing photons. Moreover, as large populations of EHB stars are relatively uncommon, it would be challenging to explain the prevalence of LIER-like emission in ETGs ($\sim30$\%, \citealt{Yan+2018a}).

We note that as an ionizing source, the term ``post-AGB'' sometimes encompasses a broader category of hot evolving stars that includes early-AGB stars, AGB-Manque stars, pre-planetary nebulae and white dwarfs \citep{Stasinska+2008}. Alternative terms for this larger category include ``hot old low mass evolved stars'' \citep[HOLMES; ][]{CidFernandes+2011} and ``hot post-horizontal branch stars,'' \citep[HPHB; ][]{Rosenfield+2012}.

There have been a few attempts to use photoionization modelling to explore the origin of LIER-like emission. Some of these studies have focused on establishing that the population of post-AGB stars in a typical ETG are capable of producing the raw number of ionizing photons required to reproduce observed \ha equivalent widths. A bulk accounting of the ionizing radiation produced by post-AGB stars indicates \ha equivalent widths of order a few angstroms \citep{Maraston+2005, CidFernandes+2011, Belfiore+2016,Gomes+2016}. Other studies have focused on reproducing emission line ratios observed in LIER-like galaxies, which is mainly sensitive to the hardness of the post-AGB star spectral energy distribution \citep[SED; e.g.,][]{Binette+1994, Stasinska+2008, Hirschmann+2017}. In \citet{Byler+2017}, we demonstrated that our nebular model was able to reproduce LIER-like emission line ratios using old stellar populations where HPHB stars are the dominant ionizing source. This work confirmed that current stellar evolution models and stellar atmospheres are capable of producing appropriately hard ionizing spectra. These studies provide useful insight into the physical origin of observed optical line ratios, but do not reveal anything about the nature of LIER-like emission or post-AGB stars as the dominant ionizing source, nor do they show simultaneous consistency with the broader emission from the larger HPHB population.

Typically, the models used for stars during the post-AGB phase are computed separately from the rest of the stars' evolution. However, in the MESA Isochrones \& Stellar Tracks \citep[MIST; ][]{Dotter+2016, Choi+2016}, stellar evolution for low-mass stars is continuously computed from the pre-main sequence phase to the end of white dwarf (WD) cooling phase. The temperatures, luminosities, and lifetimes associated with the post-AGB phase are thus a natural prediction of the stellar models. These models produce colors that are well-matched with observations throughout their evolution \citep{Choi+2016}.

In this work, we build on the work from \citet{Byler+2017}, who paired the population synthesis models from the Flexible Stellar Population Synthesis code \citep[\FSPS; ][]{Conroy+2009} with photoionization models from \Cloudy \citep{Ferland+2013} to self-consistently model the flux from galaxies. We use this model to produce the first prediction of LIER-like emission from post-AGB stars that is based on fully self-consistent stellar and photoionization modelling. As we show here, this model can simultaneously reproduce the nebular emission signatures of star forming galaxies and LIER-like emission signature observed in ETGs. We also show that this model accurately reproduces the optical properties observed in the ETG population.

We describe the stellar and nebular model in \S\ref{sec:model}. We introduce the ionizing radiation produced by hot evolved stars in \S\ref{sec:stars}, and demonstrate that our models produce LIER-like equivalent widths (\S\ref{sec:gas:ew}) and emission line ratios (\S\ref{sec:gas:ratios}). Our conclusions are summarized in \S\ref{sec:conclusions}.

\section{Description of Model}\label{sec:model}

We use the publicly available \CloudyFSPS\footnote{Read the documentation at \url{http://nell-byler.github.io/cloudyfsps/}} \citep{cloudyFSPSv1} code from \citet{Byler+2017} to compute the line and continuum emission for stellar populations from \FSPS using the photoionization code \Cloudy. The nebular model is described in detail in \citet{Byler+2017}, but we briefly summarize the process of generating model spectra in this section. We describe the stellar models in \S\ref{sec:model:stellar} and the nebular model in \S\ref{sec:model:neb}.

\subsection{The stellar model}\label{sec:model:stellar}

We generate the underlying stellar spectra using the Flexible Stellar Population Synthesis package\footnote{\url{https://github.com/cconroy20/fsps}} \citep[FSPS; ][]{Conroy+2009, Conroy+2010} via the Python interface, \pFSPS \citep{pythonFSPSdfm}. We adopt a Kroupa initial mass function \citep[IMF;][]{Kroupa+2001} with an upper and lower mass limit of 120\Msun and 0.08\Msun, respectively. We use the default parameters in \FSPS\footnote{\FSPS GitHub commit hash \texttt{3656df5}; \pFSPS GitHub commit hash \texttt{57e59f7}} unless otherwise noted.

We use evolutionary tracks from the MESA Isochrones \& Stellar Tracks \citep[MIST\footnote{Documentation, packaged model grids, and a web interpolator are available at \url{http://waps.cfa.harvard.edu/MIST/}};][]{Dotter+2016, Choi+2016}, single-star stellar evolutionary models which include the effect of stellar rotation. The evolutionary tracks are computed using the publicly available stellar evolution package Modules for Experiments in Stellar Astrophysics \citep[MESA v7503;][]{Paxton+2011,Paxton+2013, Paxton+2015} and the isochrones are generated using Aaron Dotter's publicly available \texttt{iso} package on github\footnote{available at \url{https://github.com/dotbot2000/iso}}. The MIST models cover ages from $10^5$ to $10^{10.3}$ years, initial masses from $0.1$ to $300\,$\Msun, and metallicities in the range of $-2.0 \leq$ $[\mathrm{Z}/\mathrm{H}]$ $\leq 0.5$. Abundances are solar-scaled, assuming the \citet{Asplund+2009} protosolar birth cloud bulk metallicity, for a reference solar value of $\mathrm{Z}_{\odot} = 0.0142$.  A complete description of the models can be found in \citet{Choi+2016}.

In the MIST models, stellar evolution is continuously computed from the pre-main sequence phase to the end of white dwarf (WD) cooling phase or the end of carbon burning, depending on the initial stellar mass and metallicity. We note that the post-AGB phase in the MIST models is faster and brighter than the commonly used post-AGB stellar evolution models from \citet{Vassiliadis+1994}. However, the post-AGB timescales in the MIST models are consistent with those calculated by \citet{Miller+2016} and \citet{Weiss+2009}, which are a factor of $3-10$ times shorter than older post-AGB stellar evolution models \citep{Vassiliadis+1994, Bloecker+1995}.

We are primarily concerned with the evolution of low- and intermediate-mass stars in this work, especially at late times; we briefly summarize key model parameters that affect the evolution of these stars. Mass loss for stars with initial masses below 10\Msun is treated with a combination of the \citet{Reimers+1975} prescription for the red giant branch (RGB) and the \citet{Bloecker+1995} prescription for the AGB. Both mass-loss schemes are based on global stellar properties such as the bolometric luminosity, radius, and mass:
\begin{equation}
    \dot{M}_{\mathrm{R}} = 4\times10^{-13} \eta_{\mathrm{R}} \frac{(L/L_{\odot})(R/R_{\odot})}{(M/M_{\odot})},
\end{equation}
\begin{equation}
    \dot{M}_{\mathrm{B}} = 4.83\times10^{-9} \eta_{\mathrm{B}} \frac{(L/L_{\odot})^{2.7}}{(M/M_{\odot})^{2.1}}\cdot\frac{\dot{M}_{\mathrm{R}}}{\eta_{\mathrm{R}}},
\end{equation}
in units of  $M_{\odot}\cdot\mathrm{yr}^{-1}$, and where $\eta_{\mathrm{R}}$ and $\eta_{\mathrm{B}}$ are factors of order unity. The MIST models adopt $\eta_{\mathrm{R}} = 0.1$ and $\eta_{\mathrm{B}} = 0.2$ to match the initial-final mass relation \citep{Kalirai+2009} and the AGB luminosity functions in the Magellanic Clouds \citep{Rosenfield+2014}.

\paragraph{Spectral library} We combine the MIST tracks with a new, high resolution theoretical spectral library (C3K; Conroy, Kurucz, Cargile, Castelli, in prep) based on Kurucz stellar atmosphere and spectral synthesis routines (ATLAS12 and SYNTHE). The spectra use the latest set of atomic and molecular line lists and include both lab and predicted lines. The grid was computed assuming the \citet{Asplund+2009} solar abundance scale and a constant microturbulent velocity of 2\kms.

We supplement the C3K library with alternative spectral libraries for very hot stars and stars in rapidly evolving evolutionary phases. In this work, we are primarily interested in the hot, evolved stars that are capable of producing hydrogen-ionizing radiation. The post-AGB stars use non-LTE model spectra from \citet{Rauch+2003}. We use the H-Ni composition library, which has two metallicities: solar and 10\% solar. The 10\% solar metallicity spectra are applied to populations with metallicities \logZeq{-0.5} and lower.

\subsection{The nebular model}\label{sec:model:neb}

We use the \Cloudy nebular model implemented within \FSPS to generate spectra that include nebular line and continuum emission, described in detail in \citet{Byler+2017}. The nebular model is a grid in (1) simple stellar population (SSP) age, (2) SSP and gas-phase metallicity, and (3) ionization parameter, \U, a dimensionless quantity that gives the ratio of ionizing photons to the total hydrogen density.

The default settings of the \CloudyFSPS model are most appropriate for modeling Str{\"o}mgren sphere emission in \hii regions powered by young massive stars. Extended LIER emission, however, has two significant differences we must account for. First, unlike young massive stars, we can no longer assume that the gas and ionizing stars have the same metallicity, given that the stars are much older than the surrounding gas. Second, the gas geometry is quite different. We discuss the modifications we have made to deal with these two issues.

\subsubsection{Stellar and gas phase abundances}\label{sec:model:neb:abd}

For young stellar populations, the stars and the surrounding nebular gas should have nearly identical chemical compositions. In early type galaxies, it is much more likely that the interstellar gas has experienced significant processing over the several billion years since the formation of post-AGB star progenitors. The interstellar gas has been mixed with material lost through winds from low- and intermediate-mass stars, and enriched in carbon, nitrogen and dust \citep[e.g.,][]{Griffith+2019}.
\paragraph{Fiducial gas-phase abundances} The gas phase abundances used in this work follow the \citet{Byler+2018} nebular model, which we briefly describe here. In this work, we assume that the gas is metal-rich, with a solar oxygen abundance ($12+\log_{10} (\mathrm{O}/\mathrm{H}) = 8.69$). For most elements we use the solar abundances from \citet{Grevesse+2010}, based on the results from \citet{Asplund+2009}, and adopt the depletion factors specified by \citet{Dopita+2013}. To set the relationship between N/H and O/H we use the following equation from \citet{Byler+2018}:

\begin{equation}
\begin{aligned}
    \log_{10}&(\mathrm{N}/\mathrm{O}) = \\
    & -1.5 + \log\left( 1 + e^{\frac{12 + \log_{10}(\mathrm{O}/\mathrm{H})-8.3}{0.1}}\right),
\end{aligned}
\end{equation}

and for C/H and O/H:
\begin{equation}
\begin{aligned}
    \log_{10}&(\mathrm{C}/\mathrm{O}) = \\
    & -0.8 + 0.14\cdot\left(12 + \log_{10}(\mathrm{O}/\mathrm{H})-8.0\right) \\
    & + \log\left( 1 + e^{\frac{12 + \log_{10}(\mathrm{O}/\mathrm{H})-8.0}{0.2}}\right),
\end{aligned}
\end{equation}

Both the N/H and C/H relationships with O/H were modified from the empirically calculated \citet{Dopita+2013} relationships to better match observations of star-forming galaxies below $12+\logOH = 8$.

For $12+\log_{10} (\mathrm{O}/\mathrm{H}) = 8.69$, this corresponds to $\log_{10} (\mathrm{N}/\mathrm{O}) = -1.09$ and $\log_{10} (\mathrm{N}/\mathrm{O}) = -0.26$, including the effects of dust depletion, typical of galaxies with $12 + \log \mathrm{O}/\mathrm{H} = 8.7$ \citep[e.g.,][]{Belfiore+2017b}.

Unlike in the \citet{Byler+2018} model, here, the same gas phase abundances are used regardless of the stellar metallicity. The exact elemental abundances are given in given in Table~\ref{tab:abd}.

\begin{deluxetable}{lccc}
\tablecolumns{4}
\tablecaption{Gas phase abundances and depletion factors (D) adopted for each element. The fiducial abundance set ($\rm Z_{\sun}$) has $12+\log_{10} (\mathrm{O}/\mathrm{H}) = 8.69$; the $\alpha$-enhanced abundance set (\alphaCN) has $12+\log_{10} (\mathrm{O}/\mathrm{H}) = 8.89$.}
\tablehead{
    \colhead{} &
    \colhead{$\rm Z_{\sun}$ set} &
    \colhead{\alphaCN set} &
    \colhead{}\\
        \colhead{Element} &
        \colhead{$\log_{10} (\mathrm{E} / \mathrm{H} )$} &
        \colhead{$\log_{10} (\mathrm{E} / \mathrm{H} )$} &
        \colhead{$\log_{10}( \mathrm{D} )$}
        }
\startdata
H   & 0 & 0	& 0 \\
He  & -1.01 & -1.01 & 0 \\
C   & -3.34 & -3.11 & -0.30 \\
N   & -4.42 & -3.61 & -0.05 \\
O   & -3.31 & -3.11 & -0.07 \\
Ne  & -4.07 & -3.87 & 0 \\
Na  & -5.75 & -5.75 & -1.00 \\
Mg  & -4.40 & -4.20 & -1.08 \\
Al  & -5.55 & -5.55 & -1.39 \\
Si  & -4.49 & -4.29 & -0.81 \\
S   & -4.86 & -4.66 & 0 \\
Cl  & -6.63 & -6.63 & -1.00 \\
Ar  & -5.60 & -5.60 & 0 \\
Ca  & -5.66 & -5.46 & -2.52 \\
Fe  & -4.50 & -4.50 & -1.31 \\
Ni  & -5.78 & -5.78 & -2.00 \\
\enddata
\tablecomments{Solar abundances are from \citet{Grevesse+2010} and depletion factors are from \citet{Dopita+2013}. {\it Fiducial $\rm Z_{\sun}$ set:} updated relationships for N/H and C/H with O/H from \citet{Byler+2018} correspond to $\logten (\mathrm{N}/\mathrm{O}) = -1.09$ and $\logten (\mathrm{C}/\mathrm{O}) = -0.26$, including depletion. {\it \alphaCN set:} $\alpha$-element abundances (O, Ne, Mg, Si, S, Ar, Ca) are increased by $0.2$\,dex from those in the fiducial model; C and N have been increased relative to O such that $\logten (\mathrm{N}/\mathrm{O}) = -0.48$ and $\logten (\mathrm{C}/\mathrm{O}) = -0.23$, including depletion.}\label{tab:abd}
\end{deluxetable}

\paragraph{$\alpha$-enhanced gas} We also test a second abundance set, the ``\alphaCN'' abundance set, which is enhanced in elements produced by low- and intermediate-mass stars. We modify the solar metallicity abundances from Table~\ref{tab:abd} in two distinct ways to mimick LIER-like emission conditions.

First, early-type galaxies are known to have ``$\alpha$-enhanced'' abundance ratios. Using the abundances from Table~\ref{tab:abd} as a starting point, we enhance the gas phase abundances of O, Ne, Mg, Si, S, Ar, and Ca by $+0.2\,$dex, typical for a massive elliptical galaxy with $\log \sigma = 2.3$\kms \citep[e.g.,][]{Zhu+2010, Conroy+2014, Choi+2014}.

Second, LIER-like emission may be produced in gas clouds that are quite close to the ionizing post-AGB stars, and could thus contain ejecta from AGB star winds. There is evidence that nucleosynthetic processing by low- and intermediate-mass stars does not alter the alpha abundances, but does enhance He, N and C abundances compared to typical \hii regions \citep{Karakas+2010, Maciel+2017}. We enhance C and N relative to O following \citet{Henry+2018}: $\logten (\mathrm{N}/\mathrm{O}) = -0.5$ and $\logten (\mathrm{C}/\mathrm{O}) = 0$ for the $\alpha$-enhanced $12 + \log \mathrm{O}/\mathrm{H} = 8.89$. Including the effects of dust depletion, this corresponds to $\logten (\mathrm{N}/\mathrm{O}) = -0.48$ and $\logten (\mathrm{C}/\mathrm{O}) = -0.23$. The exact ``\alphaCN'' abundances are given in Table~\ref{tab:abd}.

We note that the \alphaCN abundance set is distinct from what one would obtain using the \citet{Byler+2018} model abundance prescription at $12 + \log \mathrm{O}/\mathrm{H} = 8.89$, which would produce $\logten (\mathrm{N}/\mathrm{O}) = -0.89$ and $\logten (\mathrm{C}/\mathrm{O}) = -0.05$ including depletion (i.e., simply scaling with oxygen abundance and accounting for secondary nucleosynthetic production mechanisms for nitrogen). To reproduce the N/O ratios observed in LIER-like early-type galaxies from \citet{Yan+2018b} ($ -0.74 \lesssim \logten (\mathrm{N}/\mathrm{O}) \lesssim -0.39$) requires substantial enrichment in nitrogen relative to oxygen, roughly 0.2-0.4 dex larger than those seen in typical star-forming galaxies.

In this work we focus on gas in ETGs that is internally supplied or has undergone significant chemical processing from low- and intermediate-mass stars. There is strong evidence that the gas in a fraction of ETGs has an external origin \citep[e.g., ][]{Davis+2011, Belfiore+2017a}. There are different proposed sources for this gas, including HI accretion \citep[e.g.,][]{Oosterloo+2010} and dust accretion \citep[e.g.,][]{Rowlands+2012}, each of which would produce different predictions for elemental abundance patterns than those used here.

\subsubsection{Gas and star geometry}\label{sec:model:neb:geom}

In \citet{Byler+2017} we tested the sensitivity of emission line ratios to the assumed geometry by running models at several different ionization parameters, varying $n_{H}$ from $10-1000\,\mathrm{cm}^{-3}$, and varying the inner radius of the gas cloud from $10^{18}-10^{20}\,$cm ($0.3-30\,$pc). We found that the BPT line ratios produced by the post-AGB star models showed little sensitivity to the star-gas geometry, a result of our simplified model in which the gas exists in a plane-parallel shell surrounding a central point source of ionizing radiation. If the gas is produced by the AGB stars themselves or has a spatial distribution that differs from the distribution of stars, the geometry will differ substantially from the simplified shell used in this early work.

We modify our treatment of gas geometry by assuming the gas in early-type galaxies is much more diffuse and extended than that found in star-forming galaxies. Ionization parameters estimated from emission lines indicate much lower ionization parameters, \logUeq{-5} to \logUeq{-3.5}, indicating that the gas has lower densities and is physically further from the ionizing sources than in star-forming galaxies \citep[e.g., ][]{Binette+1994, Yan+2012}. Density estimates in early-type galaxies show typical gas densities of $1-100\,\mathrm{cm}^{-3}$ \citep[e.g., ][]{Johansson+2016, Yan+2018a}, compared to the $10-1000\,\mathrm{cm}^{-3}$ more typical of star-forming galaxies .

In this work we encompass these geometric extremes by running models with radii of $R_{\mathrm{inner}} =$ 3, 30 pc ($10^{19}$, $10^{20}$ cm) and $n_{H}=1$,10,100$\,$cm$^{-3}$. These geometries require initial stellar masses of ${\sim}10^{6-8}\Msun$ to produce ionization parameters between $10^{-5}$ and $10^{-3}$.

\begin{figure*}[ht]
  \begin{center}
    \includegraphics[width=\linewidth]{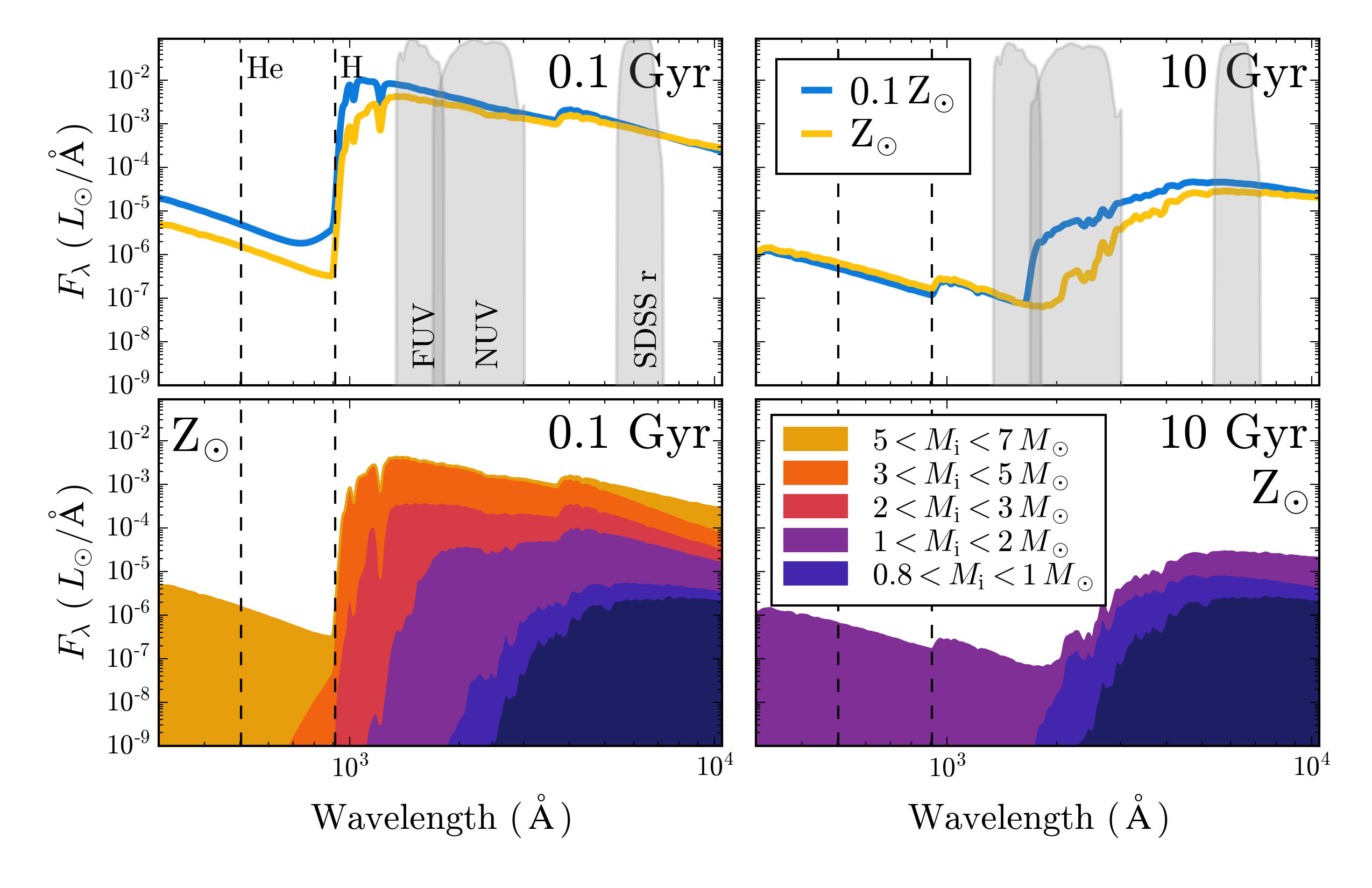}
    \caption{{\sc The UV-optical spectrum for instantaneous bursts.} \emph{Top row:} The total SED for 10\% solar (blue) and solar metallicity (yellow) populations at at 0.1\Gyr (left) and 10\Gyr (right). \emph{Bottom row:} The solar metallicity SED broken down by the relative flux contribution from stars of different initial stellar masses. In the top row, we overlay the transmission of the {\it GALEX} FUV, NUV and SDSS r-band filters. In all panels, the dashed lines indicate the ionization energy of helium ($E\geq24.6\,$eV or $\lambda \leq 504\,$\ang) and hydrogen ($E\geq13.6\,$eV or $\lambda \leq 912\,$\ang). In old stellar populations, the stars responsible for producing hydrogen-ionizing radiation are similar to the stars that dominate the UV and optical light ($1\Msun < \Mi < 2\Msun$). The spectrum blueward of the hydrogen ionization energy does not vary strongly with metallicity.}
    \label{fig:ionSpec}
  \end{center}
\end{figure*}



\section{Emission from hot evolved stars}\label{sec:stars}
\subsection{Overview}\label{sec:stars:overview}

\begin{figure*}[ht]
  \begin{center}
    \includegraphics[width=\linewidth]{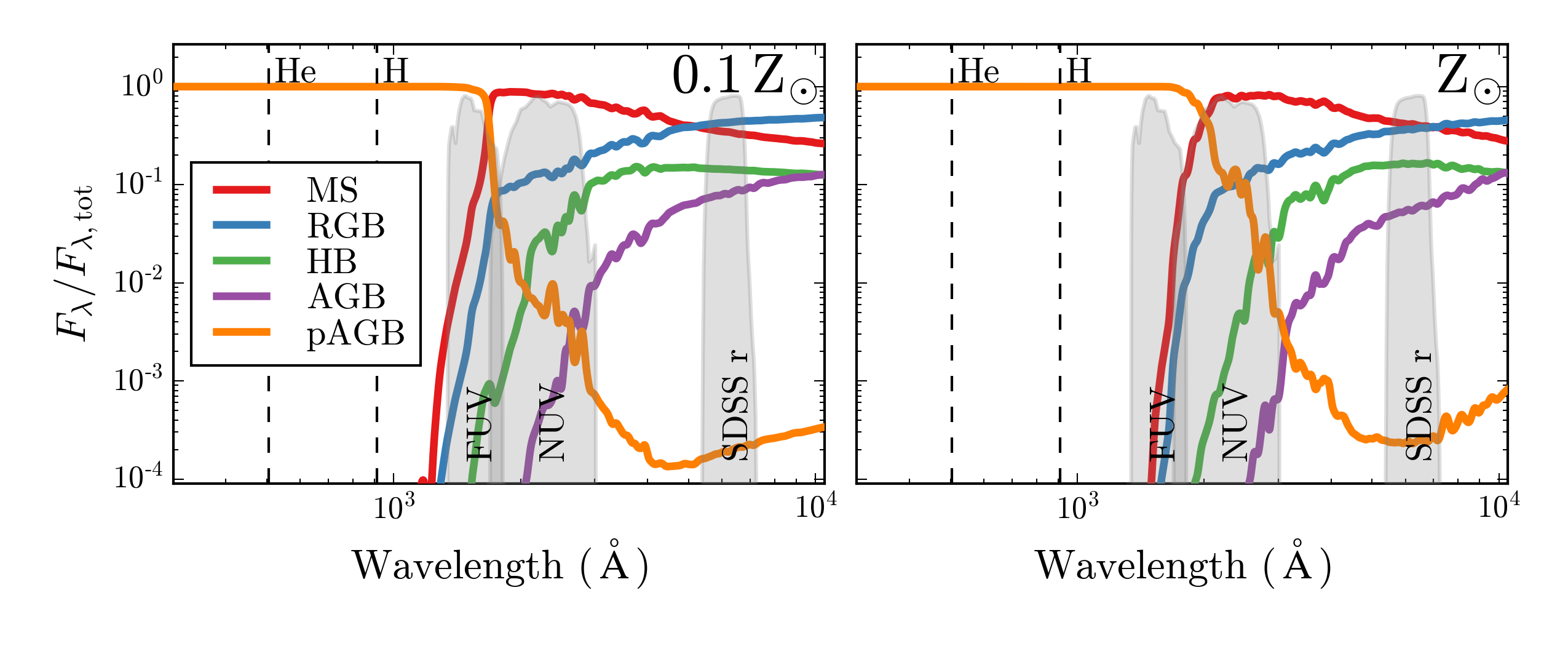}
    \caption{{\sc Flux contribution by stellar type for a 10\,Gyr SSP---} We show the fractional contribution of various evolutionary phases to the total SED at 10\% solar (left) and solar metallicity (right), including main sequence (MS), red giant branch (RGB), horizontal branch (HB), asymptotic giant branch (AGB), and post-AGB (pAGB) stars. The vertical dashed lines show the ionization energy of helium ($E\geq24.6\,$eV or $\lambda \leq 504\,$\ang) and hydrogen ($E\geq13.6\,$eV or $\lambda \leq 912\,$\ang). The grey shaded region shows the transmission of the {\it GALEX} FUV, NUV and the SDSS r-band filters.}
    \label{fig:FracSpec}
  \end{center}
\end{figure*}

In Fig.~\ref{fig:ionSpec}, we show the UV-optical spectrum produced by an old (10\Gyr) and relatively young (0.1\Gyr) instantaneous burst population. The top panel shows the total SED for solar (yellow) and 10\% solar metallicity (blue) stellar populations. Unsurprisingly, the young population is more luminous and has bluer {\it GALEX} \citep{Martin+2005} UV colors than the old population at all metallicities. For a given age, metallicity does significantly change the UV colors, with metal-poor populations producing bluer {\it GALEX} colors. However, at old ages, metallicity has very little effect on the SED blueward of the hydrogen ionization energy.

The bottom panel of Fig.~\ref{fig:ionSpec} shows the total solar metallicity SED broken down by the relative contribution from stars with different initial stellar mass, at 0.1\Gyr (\emph{left}) and 10\Gyr (\emph{right}). For the young population, stars with initial mass above 5\Msun are entirely responsible for producing photons capable of ionizing hydrogen and helium. However, those same stars contribute very little to the UV and optical flux, which are dominated by the flux from stars with initial masses between 1 and 3\Msun. In contrast, at 10\Gyr the stars that dominate the ionizing photon budget have very similar masses to the stars that dominate the UV flux, with initial masses between 1 and 2\Msun.

To understand the origin of this UV emission, Fig.~\ref{fig:FracSpec} shows the 10\Gyr spectrum broken down by stellar evolutionary type at 10\% solar (\emph{left}) and solar (\emph{right}) metallicity. We plot the fractional contribution to the total flux from main sequence (MS), red giant branch (RGB), horizontal branch (HB), AGB, and post-AGB stars, as designated by the phases tagged in the MIST isochrone tables. The AGB label includes the contribution from both early-AGB and thermally pulsing-AGB (TP-AGB) phases. We omit the WR phase from the plot, as these stars are present for $\lesssim5$\Myr in instantaneous burst populations. At both metallicities the optical light is dominated by main sequence and RGB stars.

We first consider the UV but non-ionizing flux. In general, main sequence stars still dominate the {\it GALEX} NUV band, even at 10\Gyr, while post-AGB stars dominate the FUV band flux. The populations that drive the UV colors will change with metallicity, since metallicity can change the temperature and lifetime of various stellar phases. At solar metallicity, post-AGB stars are responsible for $\sim98\%$ and $\sim10\%$ of the FUV and NUV flux, respectively. In the low metallicity model, the hotter main sequence stars contribute a higher fraction of the total flux in the {\it GALEX} FUV and NUV filters, decreasing the post-AGB contribution to $\sim90\%$ and $\sim1\%$, respectively.

\begin{figure*}[ht]
  \begin{center}
    \includegraphics[width=\linewidth]{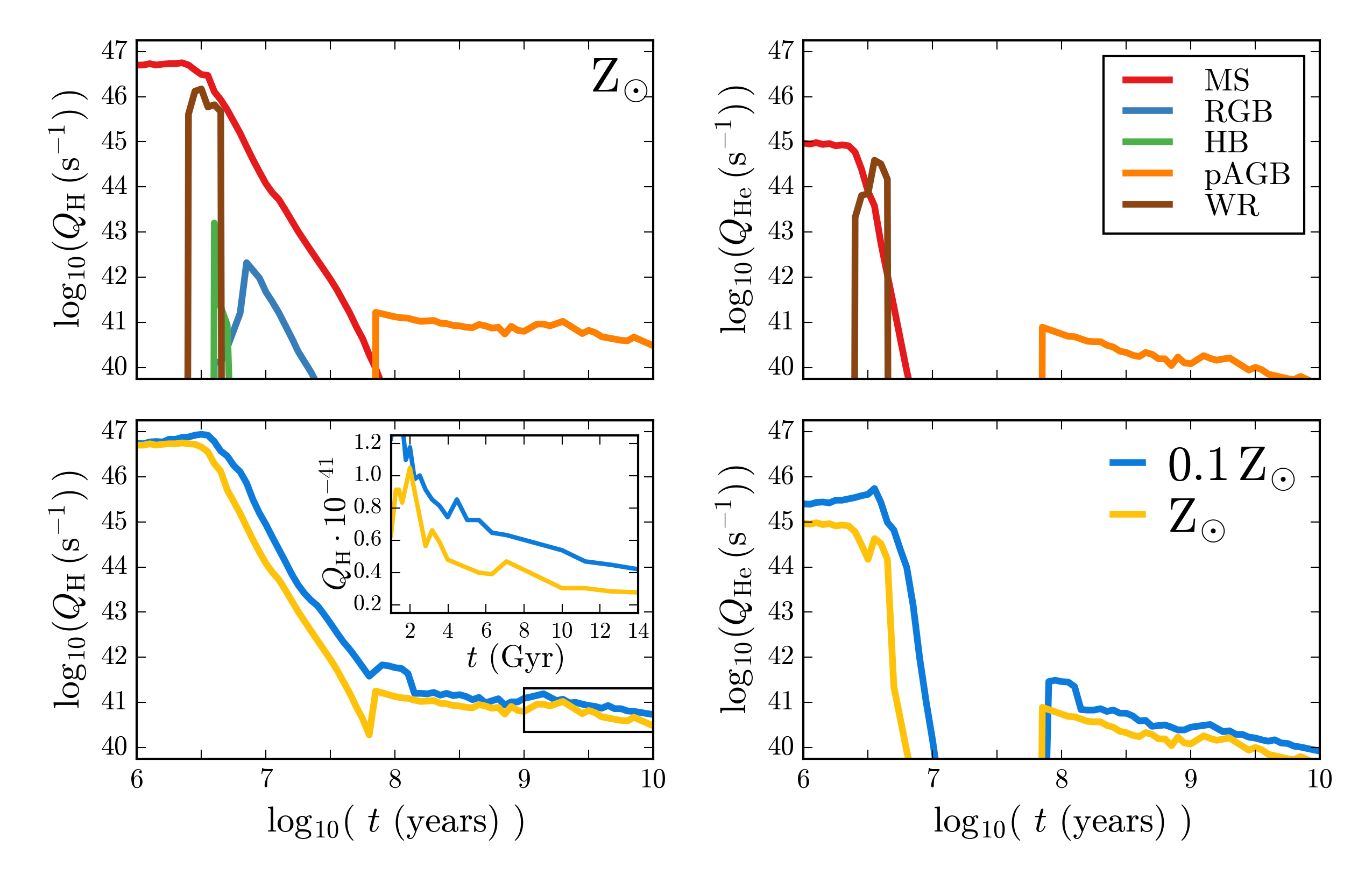}
    \caption{{\sc Time evolution of ionizing photon production---} \emph{Top row}: The ionizing photon flux per unit stellar mass for Hydrogen (\QH; \emph{left}) and Helium (\QHe, \emph{right}) for a solar-metallicity population broken down by stellar evolutionary phase. A subset of phases are shown, including Main Sequence (MS), Red Giant Branch (RGB), Horizontal Branch (HB), post-AGB (pAGB), and Wolf-Rayet (WR) stars. After the massive MS and WR stars have died, post-AGB stars provide the bulk of the ionizing radiation. These stars are hot enough to ionize both hydrogen and helium, but produce ionizing photon rates that are $10^5$ times lower than MS stars, requiring large numbers of stars to produce substantial \ha flux. Binary evolution can potentially extend the initial ionizing phase to $\sim10-20$\Myr. \emph{Bottom row}: \QH and \QHe for stellar populations at 10\% solar (blue) and solar metallicity (yellow). The inset figure in the lower left panel shows the linear evolution in \QH for old stellar populations ($t \gtrsim 1$ Gyr). For young stars, the metal-poor populations have higher ionizing photon rates and extended main-sequence lifetimes. For older populations, the evolution of \QH and \QHe is largely independent of metallicity, because the temperature evolution of post-AGB stars is an aging process rather than a metallicity-driven process.}
    \label{fig:QF}
  \end{center}
\end{figure*}

\subsection{Ionizing properties}\label{sec:stars:ion}

We now consider the ionizing flux produced by old stellar populations. As seen in Fig.~\ref{fig:FracSpec}, for a 10\Gyr SSP, the post-AGB stars are responsible for all of the flux produced at energies sufficient to ionize hydrogen and helium.

This ionizing flux evolves with time, as different stellar sources dominate the emission, as we show in the top row of Fig.~\ref{fig:QF}, broken down by stellar evolutionary type. The left column shows the time evolution of \QH, the number of photons emitted per second that are capable of ionizing hydrogen ($E\geq13.6\,$eV or $\lambda \leq 912\,$\ang) while the right column shows the time evolution of \QHe, the number of photons emitted per second that are capable of ionizing helium ($E\geq24.6\,$eV or $\lambda \leq 504\,$\ang).

For the single-star MIST models, main sequence stars dominate the ionizing photon budget for nearly 100\Myr, with the exception of a brief but intense contribution from Wolf-Rayet (WR) stars at $2-4$\Myr ($\log t\sim6.5$). The first post-AGB stars appear after a few hundred Myr. These stars are quite hot and dominate both the hydrogen- and helium-ionizing flux once they appear. At all metallicities, post-AGB stars make up more than 95\% of the ionizing flux after $\sim200$\Myr, while HB stars never contribute more than 10\%.

The bottom row of Fig.~\ref{fig:QF} shows the total ionizing flux evolution for 10\% solar (blue) and solar (yellow) populations. \QH and \QHe decline steadily and by orders of magnitude as the young, massive stars evolve off of the main sequence. \QH plateaus after ${\sim}100$\Myr as the first post-AGB stars appear, but at a level that is more than $10^5$ times smaller than the initial burst. The evolution of helium ionization is somewhat different. \QHe also declines by orders of magnitude over the first hundred million years, however, because post-AGB stars can have temperatures akin to late O-type or early B-type stars (15,000-50,000\,K), \QHe \emph{rises} after 100\Myr. The He-ionizing photon flux is still $10^5$ times smaller than during the initial burst, however.

For young, massive stars, \QH changes with metallicity, since metal line blanketing in the atmospheres of stars reduces the amount of emergent flux in the UV. In contrast, the late time evolution of \QH and \QHe is largely independent of metallicity, when post-AGB stars dominate the ionizing flux. This is a result of the fact that post-AGB models occupy a narrow range of luminosities, and the temperature evolution is driven by aging rather than by a metallicity-driven process.

\subsection{UV-optical colors}\label{sec:stars:uv}

While the optical colors of ETGs have been well characterized, the UV colors of ETGs are much more uncertain. ETGs with similar morphologies and optical colors can display a wide range of UV colors \citep[e.g.,][]{Yi+2005, Kaviraj+2007, Schawinski+2007}. Here, we briefly explore the UV properties of our models and compare them to observed ETG colors.

\begin{figure*}
  \begin{center}
    \includegraphics[width=\textwidth]{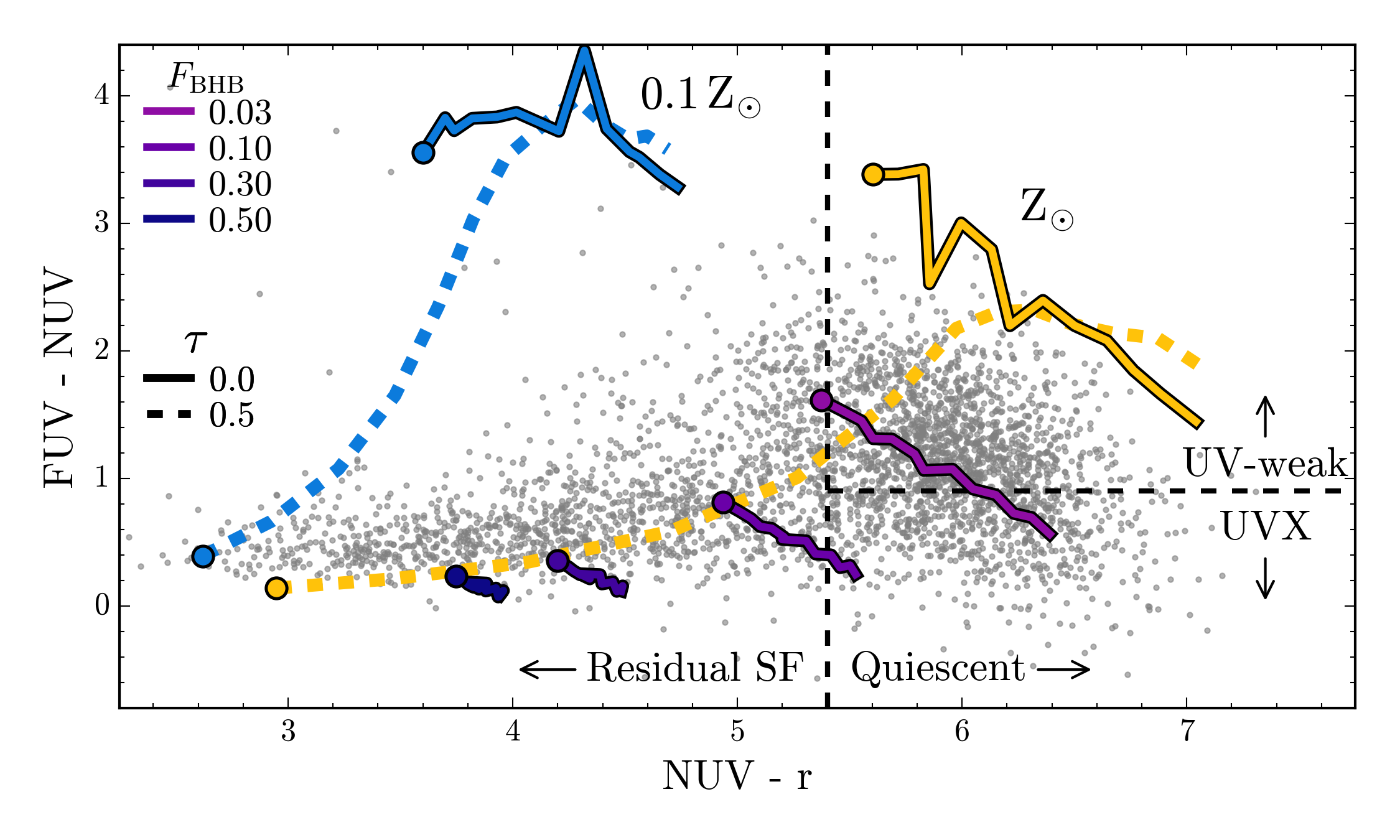}
    \caption{{\sc UV properties of hot, evolved stars---} UV-optical color-color diagram reproduced from \citet{Hernandez+2014}, with ETGs shown as grey circles. The vertical dashed line at $NUV-r=5.4$ separates red-sequence galaxies, and the horizontal line at $FUV-NUV=0.9$ shows the boundary of classic UV-upturn galaxies from \citet{Smith+2014}. We show the fiducial post-AGB star model for solar (yellow) and 10\% solar (blue) metallicities, between 3\Gyr (circle) and 14\Gyr. The solid lines represent instantaneous burst models, while the dashed lines represent models with $\tau=0.5$\Gyr. The purple lines show instantaneous burst models where the fraction of blue HB stars is modified from the fiducial model. The post-AGB stars responsible for nebular emission cannot be wholly responsible for the UV excess in some ETGs. However, even a small population of blue HB stars can produce a UV excess-like signature.}
    \label{fig:UV}
  \end{center}
\end{figure*}

Fig.~\ref{fig:UV} shows the $FUV-NUV$ \emph{\vs} $NUV-r$ color-color diagram for the $\sim3500$ ETGs from \citet{Hernandez+2014}, selected from SDSS using morphological criteria, with additional spectroscopic cuts to exclude young stellar populations and AGN. The vertical dashed line at $NUV-r=5.4$ shows the blue limit of the galaxy red sequence, where galaxies to the left of the line have some residual star formation and galaxies to the right are truly quiescent. The horizontal line at $FUV-NUV$ = 0.9 divides the red quiescent galaxies into normal, UV-weak ETGs and UV-upturn galaxies. Galaxies in the lower right quadrant are red sequence ETGs with a ``UV excess'' \citep[UVX;][]{Smith+2014}, while galaxies in the lower left are thought to have blue UV colors originating from residual star formation (RSF).

In Fig.~\ref{fig:UV} the yellow and blue lines show our fiducial model at solar and 10\% solar metallicities, for instantaneous bursts of SF (solid lines) and exponentially declining SF with $\tau=0.5$\Gyr (dashed lines) between 3 and 14\Gyr. Our fiducial models are able to reproduce the observed colors of UV-weak ETGs and ETGs with residual star formation. However, the models are not blue enough in $FUV-NUV$ to adequately reproduce the colors of ETGs with a UV excess.

Making the models bluer in $FUV-NUV$ to move them into the UVX region requires adding flux primarily to the $FUV$ filter without adding flux to the $NUV$ or $r$ bands. This change roughly corresponds to adding stars with temperatures between 20,000 and 30,000\,K, which can be achieved by simply changing the morphology of the horizontal branch stars to include BHB or EHB stars (\Teff$\geq$20,000\,K). Within \FSPS, these stars can be added with the {\tt fbhb} parameter, which takes the specified fraction of HB stars and uniformly redistributes their temperatures up to $\Teff=10^{4.5}$\,K (${\sim}$30,000$\,$K\footnote{We note that the original implementation of {\tt fbhb} in \FSPS was limited to temperatures up to $\Teff=10^{4}$\,K; we increased this to $\Teff=10^{4.5}$\,K to better represent the observed range of BHB and EHB temperatures.}).

In Fig.~\ref{fig:UV}, instantaneous burst models with $F_{\rm BHB}>0$ are shown with solid purple lines. Including a BHB population moves the models to bluer $FUV - NUV$ colors, towards the center of the observed ETG locus. Even a small fraction of BHB stars (${\sim}3$\%) can successfully move the model into the UVX portion of the diagram. However, once BHB stars dominate the $FUV$ flux ($F_{\rm BHB}\sim0.1$), the colors do not get any bluer,saturating near $FUV-NUV=0.2$. Further increases in {\tt fbhb} move the models to bluer $NUV-r$ colors at similar $FUV-NUV$ colors, masquerading as residual SF. This behavior is particular to the specific BHB implementation in \FSPS, which adopts a uniform distribution in temperature between 7,000\,K and 30,000\,K. We note that populating the UVX region of Fig.~\ref{fig:UV} requires stars with temperatures higher than 20,000\,K; BHB models with a maximum temperature of 20,000\,K do not produce enough $FUV$ flux.

In the current implementation, these hot HB stars have no discernible affect on the predicted ionizing photon budget. Within this context, the post-AGB stars produce the ionizing radiation necessary for LIER-like emission, while the hot HB stars drive the FUV and NUV behavior. We note, however, that this may not be the case for other, generative HB models \citep[e.g.,][]{Yaron+2017}.

\section{Gas ionized by hot evolved stars}\label{sec:gas}
\subsection{Emission line equivalent widths}\label{sec:gas:ew}
To measure \ha equivalent widths from our models, we largely follow the observational methodology of \citet{Belfiore+2016}, who measured \ha equivalent widths in LIER galaxies with the Mapping Nearby Galaxies at APO survey \citep[MaNGA; ][]{Bundy+2015}. In broad terms, the process involves subtracting the Balmer absorption due to the best-fit stellar continuum, and then fitting the residual emission with a Gaussian. In practice, our methodology differs from this slightly, since we are measuring model equivalent widths with a perfect knowledge of the underlying stellar continuum.

In detail, the process is as follows. We generate two spectra with \FSPS, one that includes nebular emission lines and one that does not. We use the ``non-emission'' spectrum as the best-fit stellar continuum. \FSPS returns $\mathcal{F}_{\lambda}$ in \Lsun/\ang, which we convert to cgs units (erg/s/cm$^{-2}$/\ang) by arbitrarily assuming a mass of $6 \times 10^{10}$\Msun, typical of a massive early-type galaxy, and a distance of 10$\,$Mpc. For both spectra, we scale the flux to the median flux in the continuum spectrum as measured in the 6000-6200\ang range and subtract the continuum spectrum from the emission spectrum.

We fit the resultant scaled emission-line-only spectrum with \texttt{scipy.optimize.curvefit} \citep{SciPy} using a 3-parameter Gaussian function of the form:
\begin{equation}
    f(x) = a \cdot \exp \left( \frac{-(x-b)^2}{2c^2} \right),
\end{equation}
such that the integrated flux in the Gaussian is simply $\sqrt{2\pi} \cdot ac$, an equivalent width with units of \ang.

\begin{figure*}
  \begin{center}
    \includegraphics[width=\linewidth]{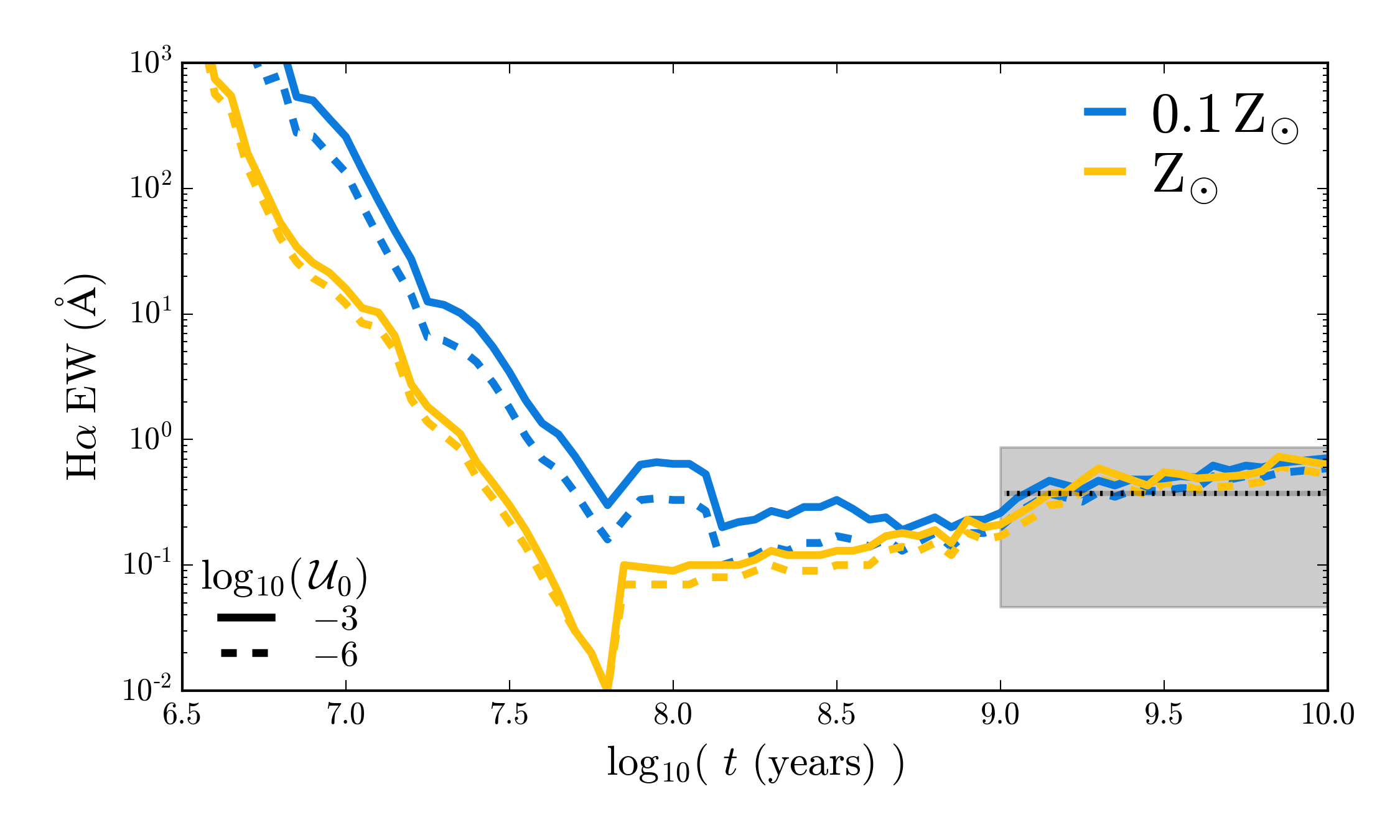}
    \caption{{\sc Time evolution of \ha equivalent widths---} Single-aged populations are shown at 10\% solar (blue) and solar metallicity (yellow). Solid lines show models with \logUeq{-3} and dashed lines show models with \logUeq{-6}. The post-AGB stars are responsible for the increase in \ha EW seen at $\log t \gtrsim 9$. The dotted line shows the median \ha EW from the \citet{Yan+2018a} sample of LIER-like red quiescent galaxies (${\sim}0.4\,$\ang), while the dark grey region shows the $25^{\mathrm{th}}$ and $75^{\mathrm{th}}$ percentiles ($0.05\,$\ang and $0.9\,$\ang, respectively).}
    \label{fig:EW}
  \end{center}
\end{figure*}

In Fig.~\ref{fig:EW} we show the \ha equivalent width as a function of time for single-aged populations after an initial burst of SF, at 10\% solar and at solar metallicity. Unsurprisingly, \ha equivalent widths are largest at young ages, when hot, young stars produce copious amounts of ionizing radiation and strong \ha emission is on top of a weaker optical stellar continuum. At the same young age, equivalent widths for metal-poor populations are larger, due to the harder ionizing spectra of metal-poor stars. The metal-poor populations also have longer main-sequence lifetimes, producing larger equivalent widths for several \Myr longer than the solar-metallicity population.

In general, the \ha equivalent widths in Fig.~\ref{fig:EW} decline with time until ${\sim}100$\Myr, when the first post-AGB stars appear. After ${\sim}1$\Gyr, equivalent widths increase as the population of post-AGB stars grows more quickly than the stellar continuum fades. The elevated \ha equivalent widths seen in the 10\% solar population just before $\log t \sim 8$ reflect the prolonged MS timescales associated with low-metallicity populations. In these populations, the final hot MS stars briefly overlap with the onset of the post-AGB phase until ${\sim}100$\Myr.

Equivalent widths measured in typical LIER-like galaxies are typically $\lesssim 3\,$\ang \citep{CidFernandes+2011}. In Fig.~\ref{fig:EW}, the dotted line shows the median \ha equivalent width from the \citet{Yan+2018a} sample of LIER-like red quiescent galaxies ($0.4\,$\ang), while the grey shaded region shows the extent of the $25^{\mathrm{th}}$ and $75^{\mathrm{th}}$ percentiles ($0.05\,$\ang and $0.9\,$\ang, respectively \citep[][\emph{private communication}]{Yan+2018a}.

At late times ($\log t \gtrsim 9$), our models produce equivalent widths consistent with observed LIERs. The solar and 10\% solar metallicity models produce equivalent widths between $0.2 - 2.5\,$\ang over the range of ionization parameters probed, with a median equivalent width of $0.5\,$\ang, in good agreement with \citet{Yan+2018a}. This is the first successful prediction for emission line equivalent widths using stellar models where the post-AGB stars have been fully modeled from the main sequence through the post-AGB phase.

\paragraph{Noise in stellar models} We note that the emission line predictions are sensitive to the hardness of the ionizing spectrum, and thus the exact spread of stellar temperatures and luminosities found in the isochrone. From Fig.~\ref{fig:EW}, at late times the equivalent widths fluctuate at the ${\sim}10\%$ level. The general behavior of the equivalent widths is quite robust, while the fluctuations are not necessarily representative of the bulk properties of the stellar population.

The fluctuations have two main sources, which we discuss in turn. The first effect is numerical, driven by interpolation amongst stellar evolution tracks during the isochrone construction process. Stochastic failures of stellar evolution tracks are caused by gaps in the grids of tracks and lead to numerical noise in the isochrones. This behavior is especially prevalent for late phases of stellar evolution, like the post-AGB phase.

The second source of fluctuations is driven by changes in the stellar physics. There is stochasticity in the TP-AGB and post-AGB phases themselves, wherein some tracks have a late thermal pulse or leave the AGB mid-pulse, which leads to different behavior in the post-AGB evolution. While not unphysical, these tracks are not representative of the bulk evolution properties and can be quite sensitive to the adopted model physics\footnote{For an in-depth discussion of this behavior, we refer the reader to Appendix A of \citet{Choi+2016}.}.

For the models shown here, we have visually inspected each isochrone and flagged those isochrones with clear numerical issues or tracks with unrepresentative evolutionary pathways. We do not include these isochrones in the final analysis. The total number of flagged isochrones is small, 14\% for the solar metallicity models. Generally, these models show a rapid and brief increase in \ha equivalent width (by as much as 1-2\,dex over $\log t = 0.05$ timescales). The inclusion of these models would skew the reported median \ha equivalent width to higher values by 0.1-0.2\ang, however we emphasize that the qualitative conclusions of this work would not change.

\begin{figure*}[ht]
  \begin{center}
    \includegraphics[width=\linewidth]{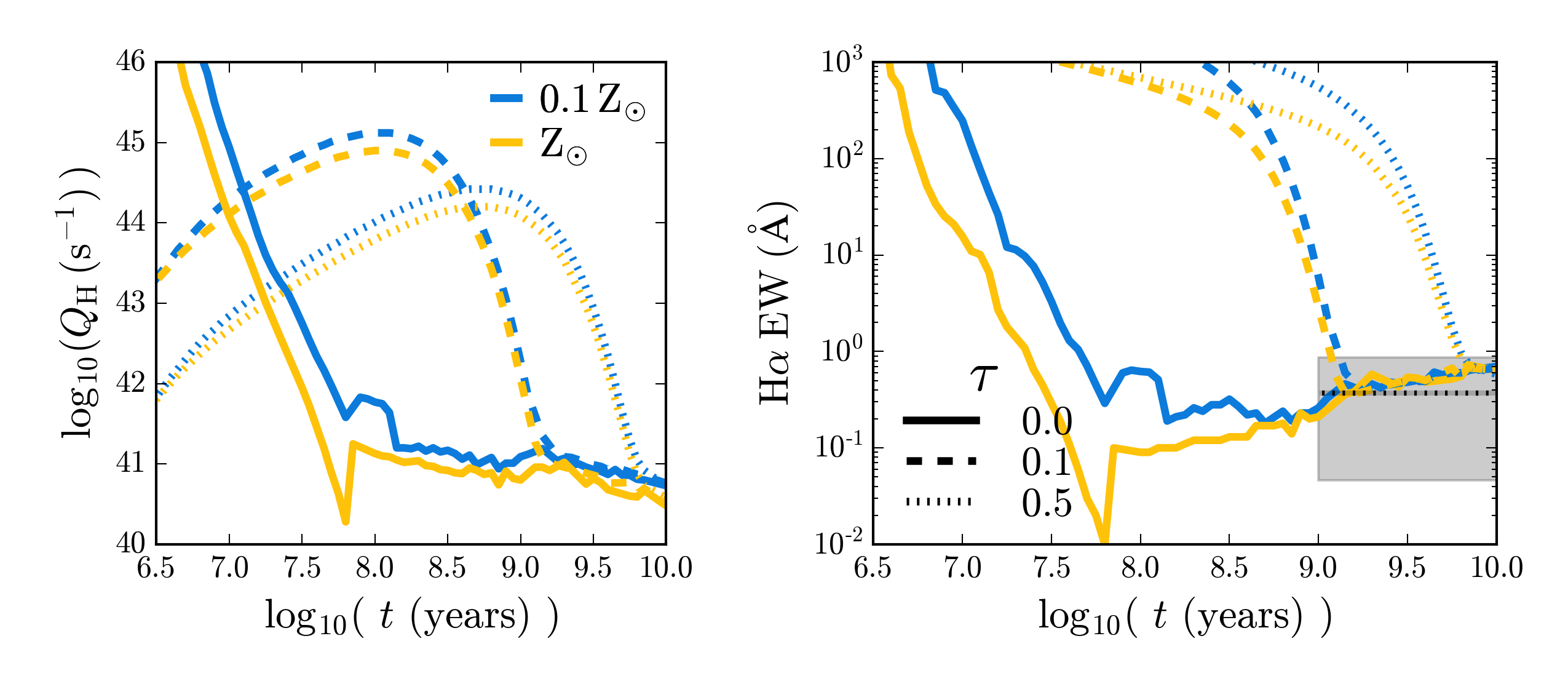}
    \caption{{\sc The dependence of \ha equivalent widths on the underlying SFH---} \emph{Left:} Time evolution of the hydrogen-ionizing photon flux per solar mass, \QH, for stellar populations with three different SFHs: $\tau = 0$ (instantaneous burst; solid line), $\tau=0.1\,$\Gyr (dashed line), and $\tau=0.5\,$\Gyr (dotted line). \emph{Right:} Time evolution of the \ha equivalent width. In both panels the 10\% solar metallicity model is shown in blue and the solar metallicity model is shown in yellow. After ${\sim}2\,$\Gyr, the $\tau = 0.1\,$\Gyr model is indistinguishable from the instantaneous burst in \QH and produces identical \ha EWs. The $\tau = 0.5\,$\Gyr produces ETG-like colors and LIER-like EWs after $5-6\,$\Gyr.
    }
    \label{fig:tau}
  \end{center}
\end{figure*}

\subsubsection{Dependence on the underlying SFH}\label{sec:gas:ew:continuum}

For young stellar populations in \hii regions, it is typical to measure \ha fluxes, since these can be tied directly to the mass of recently formed stars. For old stellar populations it is more typical to measure the flux of \ha relative to the continuum using an equivalent width (EW), which partially removes the dependence on stellar mass. The equivalent width of \ha, however, will depend on the strength of \ha \emph{and} the strength of the underlying stellar continuum at optical wavelengths.

Although it is an oversimplification to model early-type galaxies as a single-age, single-metallicity population, it is not always an inappropriate representation. To confirm that our results are robust to the choice of SFH, we extend the analysis to include more realistic SFHs, with a delayed-$\tau$ model of the form:
\begin{equation}
    \mathrm{SFR}(t) = t \exp (-t/\tau),
\end{equation}
for $\tau = $0.1 and 0.5\Gyr, and total stellar mass $10^{11}$\Msun, typical of quiescent, early-type galaxies \citep[e.g., ][]{Thomas+2005, Choi+2014}.

We compare the instantaneous burst models to the delayed-$\tau$ models in Fig.~\ref{fig:tau}. In the left panel, we show the time evolution of \QH, the ionizing photon flux. Looking first at the $\tau=0.1\,$\Gyr model, \QH remains elevated compared to an instantaneous burst, as the extended star formation continues to produce young, main sequence stars. After ${\sim}2\,$\Gyr, however, \QH evolves identically to the instantaneous burst model. Similar behavior is found in the $\tau=0.5\,$\Gyr model, but the transition to post-AGB dominated \QH occurs over longer timescales: the $\tau=0.5\,$\Gyr model follows the instantaneous burst evolution after ${\sim}5-6\,$\Gyr.

In the right panel of Fig.~\ref{fig:tau} we show the time evolution of model \ha equivalent widths. Similar to the behavior seen in \QH, the $\tau$ models follow the instantaneous burst model at late times (after approximately $2\,$\Gyr and $6\,$\Gyr for the $\tau=0.1$ and $\tau=0.5$ models, respectively).

\ha equivalent widths are proportional to the ratio of \QH, the ionizing photon flux, and inversely proportional to the flux of the underlying stellar continuum. Thus, at late times, the delayed-$\tau$ models produce \ha equivalent widths that are indistinguishable from those produced by the instantaneous burst models.

\begin{figure*}[ht]
  \begin{center}
    \includegraphics[width=\textwidth]{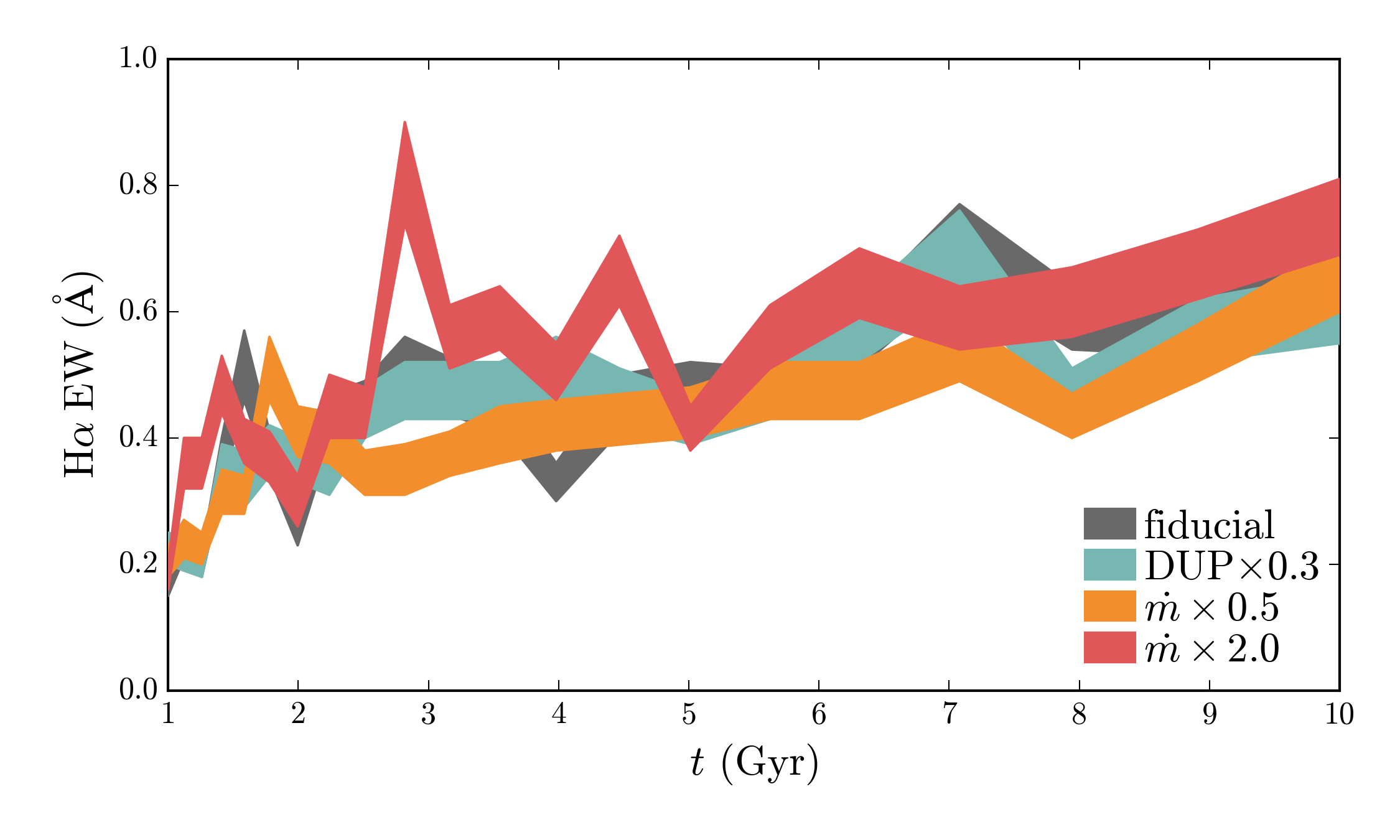}
    \caption{{\sc The dependence of \ha equivalent widths on the underlying stellar model---} The linear time evolution of \ha equivalent widths for MIST variants at solar metallicity. The filled lines represent the range of equivalent widths produced for models with \logU{} between -6 and -3. Modifying either the mass loss efficiency (orange and red) or the mixing efficiency (teal) change the timescales associated with the post-AGB phase but do not produce significant changes in \ha equivalent width.}
    \label{fig:EWvar}
  \end{center}
\end{figure*}

\subsubsection{Dependence on the underlying stellar models}\label{sec:gas:ew:isochrones}

In the MIST models, the evolution of stars at all masses is continuously computed, from the pre-main sequence phase to the end of white dwarf (WD) cooling phase. This means that we can directly probe the sensitivity of post-AGB star evolution to the various default assumptions in the evolutionary tracks.

In this section we explore modifications to the MIST models that affect the duration and intensity of the post-AGB phase to determine how this alters the resultant nebular emission. We test two variants: the overshoot mixing efficiency and mass loss efficiency. We briefly describe each of these variants in turn.

\paragraph{Overshoot mixing efficiency} In the MIST models, the convective mixing of elements in the stellar interior is implemented using the mixing length theory formalism, as described in \citet{Choi+2016}. The mixing is a time-dependent diffusive process, which is modified by overshoot mixing across convection boundaries. The method adopted by MIST follows the parametrization of \citet{Herwig+2000}, which modifies the diffusion coefficient in the overshoot region through an exponentially decaying diffusion process. The efficiency of that decay is set by $f_{ov}$, a free parameter that determines the efficiency of overshoot mixing. In this work we use MIST variants where the efficiency of overshoot mixing in the envelope of thermally pulsing (TP)-AGB stars has been increased. The default MIST model has $f_{\mathrm{ov, env}}=$0.0174. Our modified model has $f_{ov}=$0.0052, corresponding to a 30\% decrease in mixing efficiency. We refer to this model as {\tt MIST\_DUPx0.3}.

\paragraph{Mass loss efficiency} We modify the MIST mass-loss efficiency factors, $\eta_\mathrm{R}$ and $\eta_\mathrm{B}$, the Reimers and Bl{\"o}cker prescriptions for mass loss on the RGB and AGB, respectively. These prescriptions are based on global stellar properties (bolometric luminosity, radius, initial stellar mass) and are empirically calibrated to match the AGB luminosity function. In the first model variant, both $\eta_M$ and $\eta_B$ are decreased by half ({\tt MIST\_mdotx0.5}). In the second model variant, both $\eta_M$ and $\eta_B$ are doubled ({\tt MIST\_mdotx2}). We note that doubling the mass loss efficiency factors is not an unreasonable choice. The default MIST model uses $\eta_{\mathrm{R}} = 0.1$ and $\eta_{\mathrm{B}} = 0.2$, which increases to $\eta_{\mathrm{R}} = 0.2$ and $\eta_{\mathrm{B}} = 0.4$, respectively, in the {\tt MIST\_mdotx2} model. Mass loss efficiency parameters as high as 0.7 have been suggested in some cases \citep{McDonald+2015}.

We run each of the above MIST variants through \Cloudy at solar metallicity using identical input parameters. In Fig.~\ref{fig:EWvar}, we show the resultant \ha equivalent widths produced by the MIST variants as a function of time. Across all models, the predicted \ha equivalent widths vary between 0.2 and 0.9\ang. In general, the models have very similar evolution, showing \ha equivalent widths that gradually increase with time. Compared to the fiducial model, the equivalent widths from the MIST variants change by at most 30\%, and show no obvious trends with time.

There is some evidence that the {\tt MIST\_mdotx2} model produces enhanced \ha equivalent widths over the fiducial model. The variation is small, less than 30\%, and a K-S test cannot reject the hypothesis that the fiducial model and the {\tt MIST\_mdotx2} are drawn from the same distribution ($p=0.9$).

Both the mixing and mass loss efficiency affect the envelope mass for stars embarking the post-AGB phase, ultimately changing the speed at which a post-AGB star moves horizontally across the HR diagram. We had initially thought that the ionizing properties of post-AGB stars might provide additional constraints on post-AGB star timescales, which in turn could be used to assess the literature contention between post-AGB star lifetimes and the observed number counts of post-AGB stars in nearby galaxies from color-magnitude diagrams \citep[CMDs; e.g.,][]{Brown+2008, Rosenfield+2012}.

However, the lack of variation in \ha equivalent widths between the different MIST variants suggests that the ionizing photon budget in old stellar populations is dominated by \emph{proto-WD} post-AGB stars. Once the post-AGB star reaches its hottest point and turns down to the WD cooling track, timescales are much longer and are instead are governed by the physics of energy loss from the proto-WD. The properties considered here (mass loss efficiency, mixing efficiency) are sensitive to post-AGB CMD-crossing timescales rather than proto-WD cooling track timescales, and are thus unlikely to be connected to the LIER-like emission produced by post-AGB stars.

\begin{figure*}[ht]
  \begin{center}
    \includegraphics[width=\linewidth]{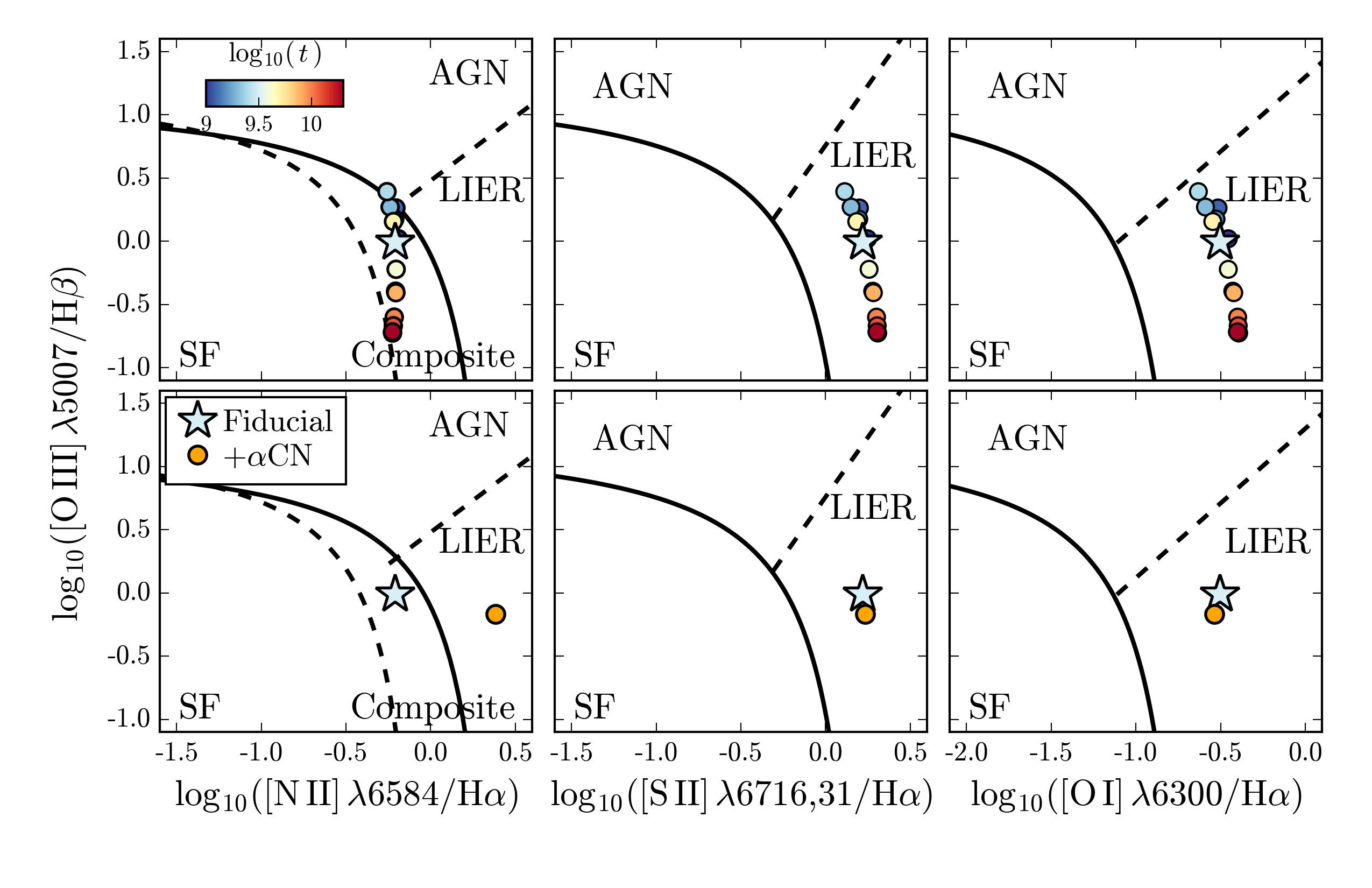}
    \caption{{\sc LIER-like emission line ratios from post-AGB stars---} Diagnostic diagrams from left to right: \nii/\ha \vs \oiii/\hb, \sii/\ha \vs \oiii/\hb, and  \oi/\ha \vs \oiii/\hb. \emph{Top row:} Line ratios from the fiducial model, assuming a solar metallicity ionizing spectrum, solar metallicity gas, and $n_{\mathrm{H}}=10$\,cm$^{-3}$. Models between 1 and 14\Gyr are shown, with markers color-coded by age. \emph{Bottom Row:} Line ratio variations driven by changes gas phase abundances. The fiducial model at 10\Gyr is shown with a blue star. The orange circle shows the same model run with the $\alpha-$element-, nitrogen-, and carbon-enhanced ``\alphaCN'' abundance set. We include lines from the literature used to separate emission driven by star formation (SF), active galactic nuclei (AGN), and low-ionization emission regions (LIERs), from \citet{Kewley+2001}, \citet{Kauffmann+2003b}, \citet{Kewley+2006}, and \citet{CidFernandes+2010} as described in the text. In the middle and right diagnostic diagrams, the post-AGB models produce line ratios consistent with LIER-like emission. In the standard BPT diagram (left), only models with N-enhanced gas reproduce LIER-like line ratios.}
    \label{fig:BPT}
  \end{center}
\end{figure*}

\subsection{Emission line ratios}\label{sec:gas:ratios}

In Fig.~\ref{fig:BPT} we show several emission line diagnostic diagrams that are commonly used to separate objects with different ionizing spectra. The standard Baldwin, Phillips, \& Terlevich \citep[BPT;][]{BPT} diagram is shown in the left column, which uses the \nii/\ha and \oiii/\hb line ratios. The center and right columns show the \sii/\ha \vs \oiii/\hb and \oi/\ha \vs \oiii/\hb diagrams, respectively. The latter two diagnostic diagrams have been highlighted as sensitive LIER diagnostics, since they make use of low-ionization lines. We include empirically derived relationships used to separate objects with different ionizing sources. In the BPT diagram (left panel), the dashed line shows the \citet{Kauffmann+2003b} separation between SF and composite regions, the solid black line shows the \citet{Kewley+2001} separation between AGN and SF, and the diagonal dashed line shows the \citet{CidFernandes+2010} translation of the \citet{Kewley+2006} criteria to separate AGN and LIERs. In the center and right diagrams, we show the \citet{Kewley+2006} criteria used to separate SF, AGN, and LIERs.

The top row of Fig.~\ref{fig:BPT} shows line ratios from the fiducial post-AGB star model, which uses a solar metallicity stellar population ionizing spectrum, standard solar gas phase abundances, and a gas phase density of $n_{\mathrm{H}}=10$ cm$^{-3}$. For visual clarity, we do not show results from the low metallicity spectrum, because the ionizing spectra have very similar hardness (see Fig.~\ref{fig:ionSpec}), producing gas with similar ionization states and thus similar emission line ratios. Each marker is color-coded by the age of the stellar population, which varies between 1 and 14\Gyr (blue to red).

Fig.~\ref{fig:BPT} shows that the models produce a wide range of \oiii/\hb ratios for a comparatively narrow range in \nii/\ha. The oldest models produce the lowest \oiii/\hb ratios, reflecting the decrease in \QH with time (Fig.~\ref{fig:QF}).

In each of the diagnostic diagrams shown in the top row of Fig.~\ref{fig:BPT}, the fiducial post-AGB star models occupy a region distinct from star forming galaxies. In the middle and right panels, which exclusively use low ionization emission lines, the predicted line ratios are consistent with line ratios observed in LIER-like galaxies. In the traditional BPT diagram (upper left), the model \nii/\ha line ratios occupy the so-called ``composite'' region, with ratios that are larger than those observed in SF galaxies. Unlike the low-ionization diagnostics shown in the middle and right panels, the model line ratios do not fall in the LIER region of the classic BPT diagram. For the range of ionization parameters, gas densities, radii, and stellar ages considered in this work (\S~\ref{sec:model:neb}), there is no combination of model parameters that adequately populates the LIER region of the BPT diagram, suggesting that populating this region would require a change to the fiducial model.

We explore this possibility in the bottom row of Fig.~\ref{fig:BPT}, which shows line ratio variations driven by changes to the fiducial gas phase abundances for a fixed age of 3\Gyr. The fiducial model (shown with the blue star) assumes a solar metallicity stellar population and solar-like gas phase abundances, with $n_{\mathrm{H}}=10$\,cm$^{-3}$. The blue circle shows the fiducial model run through \Cloudy with the $\alpha$-, nitrogen-, and carbon-enhanced ``\alphaCN'' abundance set (\S\ref{sec:model:neb:abd}).

The \alphaCN model shows the biggest shift in line ratios in the classic BPT diagram, where the enhanced nitrogen abundance increases \nii/\ha by nearly 0.5\,dex, pushing the line ratios well into the LIER region of the diagram. The enhanced \alphaCN model otherwise produces only second-order effects on other emission line ratios. The \alphaCN models show slightly lower \oiii/\hb ratios, simply due to the cooler nebulae temperatures produced by the increased oxygen abundance. This suggests that ETGs with LIER-like emission have N/O ratios 0.2-0.6\,dex larger than what is typically observed in star-forming galaxies, consistent with the findings of \citet{Yan+2018a}.



\section{Conclusions}\label{sec:conclusions}

In this work, we present the first prediction of LIER-like emission from post-AGB stars that is based on fully self-consistent stellar models and photoionization modelling. We use these models to characterize the physical origin of LIER-like emission and post-AGB stars as an ionizing source. We summarize our conclusions below.

\begin{itemize}
    \item For instantaneous bursts, post-AGB stars make up more than 95\% of the ionizing flux after ${\sim}100$\Myr, while horizontal branch stars never contribute more than 10\% of the total ionizing flux, assuming the default \FSPS parameters (\S\ref{sec:stars:ion}).
    \item Post-AGB star models produce \ha equivalent widths between 0.1 and 2.5\ang. Equivalent widths increase with age, driven primarily by the dimming of the $r$-band continuum as the underlying stellar population ages (\S\ref{sec:gas:ew}).
    \item Post-AGB star models produce emission line ratios distinct from young stellar populations in several emission line diagnostic diagrams. The post-AGB model produces line ratios in the LIER region of the \sii/\ha and \oi/\ha diagrams, and in the ``composite'' region of the standard BPT diagram (\S\ref{sec:gas:ratios}).
    \item In the standard BPT diagram, models with enhanced $\alpha$, C, and N gas phase abundances produce LIER-like line ratios, due to the increased nitrogen abundance. This supports the idea that the gas in LIER-like galaxies is composed of AGB star ejecta and/or has been substantially enriched by low- and intermediate-mass stars (\S\ref{sec:gas:ratios}).
    \item Post-AGB stars have very similar ionizing spectra at all metallicities. The similar hardness in the ionizing spectrum produces gas with similar ionization states, thus, emission line ratios from post-AGB stars do not vary much with stellar metallicity. The metallicity of the stellar population does, however, change the measured equivalent widths, since the underlying stellar continuum varies with metallicity, as does the absolute number of ionizing photons produced per solar mass (\S\ref{sec:gas:ew}).
    \item We have tested the sensitivity of the post-AGB phase to several of the default parameters used in the MIST models, including the mass loss efficiency and the overshoot mixing efficiency. Generally, these parameters have only a small effect on the bulk ionizing properties of post-AGB stars. Increased mass loss efficiency increases the resultant \ha equivalent widths, but only by 10-30\%. The lack of variation in \ha equivalent widths between the different MIST variants suggests that the ionizing photon budget in old stellar populations is dominated by proto-WD post-AGB stars (\S\ref{sec:gas:ew:isochrones}).
    \item Post-AGB stars as implemented in MIST are capable of driving LIER-like emission, and contribute to the UV-excess observed in ETGs. For the set of models presented here, we find that the UVX region of the $FUV-NUV$ versus $NUV-r$ diagram can only be populated by including a small population of blue HB stars. These stars do not contribute significantly to the ionizing photon budget, but drive substantial changes in the UV flux (\S\ref{sec:stars:uv}).
\end{itemize}

We have demonstrated that the self-consistent stellar models for post-AGB stars have optical properties consistent with ETG galaxies and produce LIER-like emission. The UV colors of ETGs present a much more puzzling picture, however. We plan to study the detailed UV properties of galaxies with LIER-like emission in more detail in future work.

\acknowledgments

Special thanks to Renbin Yan, for sharing with us detailed statistics from the \citet{Yan+2018a} sample of LIER-like ETGs. We would also like to thank the anonymous referee for thorough and constructive feedback that greatly improved this work. C.C. acknowledges support from the Packard Foundation. N.B. acknowledges support from the University of Washington's Royalty Research Fund Grant 65-8055, and the Australian Research Council Centre of Excellence for All Sky Astrophysics in 3 Dimensions (ASTRO 3D), through project number CE170100013. Some of this material is based upon work supported by the National Science Foundation under Award No. 1501205. This research has made use of NASA's Astrophysics Data System.

\software{\Cloudy$\;\mathrm{v}13.03$ \citep{Ferland+2013},
          {\tt FSPS}$\;\mathrm{v}3.0$ \citep{Conroy+2009, Conroy+2010},
          {\tt cloudyFSPS}$\;\mathrm{v}1.0$ \citep{cloudyFSPSv1},
          {\tt pythonFSPS}$\;\mathrm{v}0.1.1$ \citep{pythonFSPSdfm}
         }
\bibliographystyle{aasjournal}
\bibliography{main}

\begin{thebibliography}{}
\expandafter\ifx\csname natexlab\endcsname\relax\def\natexlab#1{#1}\fi
\providecommand{\url}[1]{\href{#1}{#1}}
\providecommand{\dodoi}[1]{doi:~\href{http://doi.org/#1}{\nolinkurl{#1}}}
\providecommand{\doeprint}[1]{\href{http://ascl.net/#1}{\nolinkurl{http://ascl.net/#1}}}
\providecommand{\doarXiv}[1]{\href{https://arxiv.org/abs/#1}{\nolinkurl{https://arxiv.org/abs/#1}}}

\bibitem[{{Allen} {et~al.}(2008){Allen}, {Groves}, {Dopita}, {Sutherland}, \&
  {Kewley}}]{Allen+2008}
{Allen}, M.~G., {Groves}, B.~A., {Dopita}, M.~A., {Sutherland}, R.~S., \&
  {Kewley}, L.~J. 2008, \apjs, 178, 20, \dodoi{10.1086/589652}

\bibitem[{{Asplund} {et~al.}(2009){Asplund}, {Grevesse}, {Sauval}, \&
  {Scott}}]{Asplund+2009}
{Asplund}, M., {Grevesse}, N., {Sauval}, A.~J., \& {Scott}, P. 2009, \araa, 47,
  481, \dodoi{10.1146/annurev.astro.46.060407.145222}

\bibitem[{{Baldwin} {et~al.}(1981){Baldwin}, {Phillips}, \& {Terlevich}}]{BPT}
{Baldwin}, J.~A., {Phillips}, M.~M., \& {Terlevich}, R. 1981, \pasp, 93, 5,
  \dodoi{10.1086/130766}

\bibitem[{{Belfiore} {et~al.}(2016){Belfiore}, {Maiolino}, {Maraston},
  {Emsellem}, {Bershady}, {Masters}, {Yan}, {Bizyaev}, {Boquien}, {Brownstein},
  {Bundy}, {Drory}, {Heckman}, {Law}, {Roman-Lopes}, {Pan}, {Stanghellini},
  {Thomas}, {Weijmans}, \& {Westfall}}]{Belfiore+2016}
{Belfiore}, F., {Maiolino}, R., {Maraston}, C., {et~al.} 2016, \mnras, 461,
  3111, \dodoi{10.1093/mnras/stw1234}

\bibitem[{{Belfiore} {et~al.}(2017{\natexlab{a}}){Belfiore}, {Maiolino},
  {Tremonti}, {S{\'a}nchez}, {Bundy}, {Bershady}, {Westfall}, {Lin}, {Drory},
  {Boquien}, {Thomas}, \& {Brinkmann}}]{Belfiore+2017b}
{Belfiore}, F., {Maiolino}, R., {Tremonti}, C., {et~al.} 2017{\natexlab{a}},
  \mnras, 469, 151, \dodoi{10.1093/mnras/stx789}

\bibitem[{{Belfiore} {et~al.}(2017{\natexlab{b}}){Belfiore}, {Maiolino},
  {Maraston}, {Emsellem}, {Bershady}, {Masters}, {Bizyaev}, {Boquien},
  {Brownstein}, {Bundy}, {Diamond-Stanic}, {Drory}, {Heckman}, {Law},
  {Malanushenko}, {Oravetz}, {Pan}, {Roman-Lopes}, {Thomas}, {Weijmans},
  {Westfall}, \& {Yan}}]{Belfiore+2017a}
{Belfiore}, F., {Maiolino}, R., {Maraston}, C., {et~al.} 2017{\natexlab{b}},
  \mnras, 466, 2570, \dodoi{10.1093/mnras/stw3211}

\bibitem[{{Binette} {et~al.}(1994){Binette}, {Magris}, {Stasi{\'n}ska}, \&
  {Bruzual}}]{Binette+1994}
{Binette}, L., {Magris}, C.~G., {Stasi{\'n}ska}, G., \& {Bruzual}, A.~G. 1994,
  \aap, 292, 13

\bibitem[{{Bloecker}(1995)}]{Bloecker+1995}
{Bloecker}, T. 1995, \aap, 299, 755

\bibitem[{{Bressan} {et~al.}(2012){Bressan}, {Marigo}, {Girardi}, {Salasnich},
  {Dal Cero}, {Rubele}, \& {Nanni}}]{Bressan+2012}
{Bressan}, A., {Marigo}, P., {Girardi}, L., {et~al.} 2012, \mnras, 427, 127,
  \dodoi{10.1111/j.1365-2966.2012.21948.x}

\bibitem[{{Brown} {et~al.}(2008){Brown}, {Smith}, {Ferguson}, {Sweigart},
  {Kimble}, \& {Bowers}}]{Brown+2008}
{Brown}, T.~M., {Smith}, E., {Ferguson}, H.~C., {et~al.} 2008, \apj, 682, 319,
  \dodoi{10.1086/589611}

\bibitem[{{Bundy} {et~al.}(2015){Bundy}, {Bershady}, {Law}, {Yan}, {Drory},
  {MacDonald}, {Wake}, {Cherinka}, {S{\'a}nchez-Gallego}, {Weijmans}, {Thomas},
  {Tremonti}, {Masters}, {Coccato}, {Diamond-Stanic}, {Arag{\'o}n-Salamanca},
  {Avila-Reese}, {Badenes}, {Falc{\'o}n-Barroso}, {Belfiore}, {Bizyaev},
  {Blanc}, {Bland-Hawthorn}, {Blanton}, {Brownstein}, {Byler}, {Cappellari},
  {Conroy}, {Dutton}, {Emsellem}, {Etherington}, {Frinchaboy}, {Fu}, {Gunn},
  {Harding}, {Johnston}, {Kauffmann}, {Kinemuchi}, {Klaene}, {Knapen},
  {Leauthaud}, {Li}, {Lin}, {Maiolino}, {Malanushenko}, {Malanushenko}, {Mao},
  {Maraston}, {McDermid}, {Merrifield}, {Nichol}, {Oravetz}, {Pan}, {Parejko},
  {Sanchez}, {Schlegel}, {Simmons}, {Steele}, {Steinmetz}, {Thanjavur},
  {Thompson}, {Tinker}, {van den Bosch}, {Westfall}, {Wilkinson}, {Wright},
  {Xiao}, \& {Zhang}}]{Bundy+2015}
{Bundy}, K., {Bershady}, M.~A., {Law}, D.~R., {et~al.} 2015, \apj, 798, 7,
  \dodoi{10.1088/0004-637X/798/1/7}

\bibitem[{{Byler}(2018)}]{cloudyFSPSv1}
{Byler}, N. 2018, cloudyFSPS, 1.0.0,  Zenodo, \dodoi{10.5281/zenodo.1156412}.
\newblock \url{https://doi.org/10.5281/zenodo.1156412}

\bibitem[{{Byler} {et~al.}(2017){Byler}, {Dalcanton}, {Conroy}, \&
  {Johnson}}]{Byler+2017}
{Byler}, N., {Dalcanton}, J.~J., {Conroy}, C., \& {Johnson}, B.~D. 2017, \apj,
  840, 44, \dodoi{10.3847/1538-4357/aa6c66}

\bibitem[{{Byler} {et~al.}(2018){Byler}, {Dalcanton}, {Conroy}, {Johnson},
  {Levesque}, \& {Berg}}]{Byler+2018}
{Byler}, N., {Dalcanton}, J.~J., {Conroy}, C., {et~al.} 2018, \apj, 863, 14,
  \dodoi{10.3847/1538-4357/aacd50}

\bibitem[{{Catelan}(2009)}]{Catelan+2009}
{Catelan}, M. 2009, \apss, 320, 261, \dodoi{10.1007/s10509-009-9987-8}

\bibitem[{{Choi} {et~al.}(2014){Choi}, {Conroy}, {Moustakas}, {Graves},
  {Holden}, {Brodwin}, {Brown}, \& {van Dokkum}}]{Choi+2014}
{Choi}, J., {Conroy}, C., {Moustakas}, J., {et~al.} 2014, \apj, 792, 95,
  \dodoi{10.1088/0004-637X/792/2/95}

\bibitem[{{Choi} {et~al.}(2016){Choi}, {Dotter}, {Conroy}, {Cantiello},
  {Paxton}, \& {Johnson}}]{Choi+2016}
{Choi}, J., {Dotter}, A., {Conroy}, C., {et~al.} 2016, \apj, 823, 102,
  \dodoi{10.3847/0004-637X/823/2/102}

\bibitem[{{Cid Fernandes} {et~al.}(2011){Cid Fernandes}, {Stasi{\'n}ska},
  {Mateus}, \& {Vale Asari}}]{CidFernandes+2011}
{Cid Fernandes}, R., {Stasi{\'n}ska}, G., {Mateus}, A., \& {Vale Asari}, N.
  2011, \mnras, 413, 1687, \dodoi{10.1111/j.1365-2966.2011.18244.x}

\bibitem[{{Cid Fernandes} {et~al.}(2010){Cid Fernandes}, {Stasi{\'n}ska},
  {Schlickmann}, {Mateus}, {Vale Asari}, {Schoenell}, \&
  {Sodr{\'e}}}]{CidFernandes+2010}
{Cid Fernandes}, R., {Stasi{\'n}ska}, G., {Schlickmann}, M.~S., {et~al.} 2010,
  \mnras, 403, 1036, \dodoi{10.1111/j.1365-2966.2009.16185.x}

\bibitem[{{Conroy} {et~al.}(2014){Conroy}, {Graves}, \& {van
  Dokkum}}]{Conroy+2014}
{Conroy}, C., {Graves}, G.~J., \& {van Dokkum}, P.~G. 2014, \apj, 780, 33,
  \dodoi{10.1088/0004-637X/780/1/33}

\bibitem[{{Conroy} \& {Gunn}(2010)}]{Conroy+2010}
{Conroy}, C., \& {Gunn}, J.~E. 2010, \apj, 712, 833,
  \dodoi{10.1088/0004-637X/712/2/833}

\bibitem[{{Conroy} {et~al.}(2009){Conroy}, {Gunn}, \& {White}}]{Conroy+2009}
{Conroy}, C., {Gunn}, J.~E., \& {White}, M. 2009, \apj, 699, 486,
  \dodoi{10.1088/0004-637X/699/1/486}

\bibitem[{{Davis} {et~al.}(2011){Davis}, {Alatalo}, {Sarzi}, {Bureau}, {Young},
  {Blitz}, {Serra}, {Crocker}, {Krajnovi{\'c}}, {McDermid}, {Bois}, {Bournaud},
  {Cappellari}, {Davies}, {Duc}, {de Zeeuw}, {Emsellem}, {Khochfar},
  {Kuntschner}, {Lablanche}, {Morganti}, {Naab}, {Oosterloo}, {Scott}, \&
  {Weijmans}}]{Davis+2011}
{Davis}, T.~A., {Alatalo}, K., {Sarzi}, M., {et~al.} 2011, \mnras, 417, 882,
  \dodoi{10.1111/j.1365-2966.2011.19355.x}

\bibitem[{{Dopita} \& {Sutherland}(1995)}]{Dopita+1995}
{Dopita}, M.~A., \& {Sutherland}, R.~S. 1995, \apj, 455, 468,
  \dodoi{10.1086/176596}

\bibitem[{{Dopita} {et~al.}(2013){Dopita}, {Sutherland}, {Nicholls}, {Kewley},
  \& {Vogt}}]{Dopita+2013}
{Dopita}, M.~A., {Sutherland}, R.~S., {Nicholls}, D.~C., {Kewley}, L.~J., \&
  {Vogt}, F.~P.~A. 2013, \apjs, 208, 10, \dodoi{10.1088/0067-0049/208/1/10}

\bibitem[{{Dotter}(2016)}]{Dotter+2016}
{Dotter}, A. 2016, \apjs, 222, 8, \dodoi{10.3847/0067-0049/222/1/8}

\bibitem[{{Ferland} \& {Netzer}(1983)}]{Ferland+1983}
{Ferland}, G.~J., \& {Netzer}, H. 1983, \apj, 264, 105, \dodoi{10.1086/160577}

\bibitem[{{Ferland} {et~al.}(2013){Ferland}, {Porter}, {van Hoof}, {Williams},
  {Abel}, {Lykins}, {Shaw}, {Henney}, \& {Stancil}}]{Ferland+2013}
{Ferland}, G.~J., {Porter}, R.~L., {van Hoof}, P.~A.~M., {et~al.} 2013, \rmxaa,
  49, 137.
\newblock \doarXiv{1302.4485}

\bibitem[{{Filippenko}(2003)}]{Filippenko+2003}
{Filippenko}, A.~V. 2003, in Astronomical Society of the Pacific Conference
  Series, Vol. 290, Active Galactic Nuclei: From Central Engine to Host Galaxy,
  ed. S.~{Collin}, F.~{Combes}, \& I.~{Shlosman}, 369

\bibitem[{{Foreman-Mackey} {et~al.}(2014){Foreman-Mackey}, {Sick}, \&
  {Johnson}}]{pythonFSPSdfm}
{Foreman-Mackey}, D., {Sick}, J., \& {Johnson}, B.~D. 2014, python-fsps: Python
  bindings to FSPS, 0.1.1,  Zenodo, \dodoi{10.5281/zenodo.12157}.
\newblock \url{https://doi.org/10.5281/zenodo.12157}

\bibitem[{{Gomes} {et~al.}(2016){Gomes}, {Papaderos}, {Kehrig},
  {V{\'{\i}}lchez}, {Lehnert}, {S{\'a}nchez}, {Ziegler}, {Breda}, {Dos Reis},
  {Iglesias-P{\'a}ramo}, {Bland-Hawthorn}, {Galbany}, {Bomans},
  {Rosales-Ortega}, {Cid Fernandes}, {Walcher}, {Falc{\'o}n-Barroso},
  {Garc{\'{\i}}a-Benito}, {M{\'a}rquez}, {Del Olmo}, {Masegosa}, {Moll{\'a}},
  {Marino}, {Gonz{\'a}lez Delgado}, {L{\'o}pez-S{\'a}nchez}, \& {CALIFA
  Collaboration}}]{Gomes+2016}
{Gomes}, J.~M., {Papaderos}, P., {Kehrig}, C., {et~al.} 2016, \aap, 588, A68,
  \dodoi{10.1051/0004-6361/201525976}

\bibitem[{{Goudfrooij} {et~al.}(1994){Goudfrooij}, {Hansen}, {Jorgensen}, \&
  {Norgaard-Nielsen}}]{Goudfrooij+1994}
{Goudfrooij}, P., {Hansen}, L., {Jorgensen}, H.~E., \& {Norgaard-Nielsen},
  H.~U. 1994, \aaps, 105, 341

\bibitem[{{Grevesse} {et~al.}(2010){Grevesse}, {Asplund}, {Sauval}, \&
  {Scott}}]{Grevesse+2010}
{Grevesse}, N., {Asplund}, M., {Sauval}, A.~J., \& {Scott}, P. 2010, \apss,
  328, 179, \dodoi{10.1007/s10509-010-0288-z}

\bibitem[{{Griffith} {et~al.}(2019){Griffith}, {Martini}, \&
  {Conroy}}]{Griffith+2019}
{Griffith}, E., {Martini}, P., \& {Conroy}, C. 2019, \mnras, 484, 562,
  \dodoi{10.1093/mnras/sty3405}

\bibitem[{{Halpern} \& {Steiner}(1983)}]{Halpern+1983}
{Halpern}, J.~P., \& {Steiner}, J.~E. 1983, \apjl, 269, L37,
  \dodoi{10.1086/184051}

\bibitem[{{Heckman}(1981)}]{Heckman+1981}
{Heckman}, T.~M. 1981, \apjl, 250, L59, \dodoi{10.1086/183674}

\bibitem[{{Heckman} {et~al.}(1998){Heckman}, {Robert}, {Leitherer}, {Garnett},
  \& {van der Rydt}}]{Heckman+1998}
{Heckman}, T.~M., {Robert}, C., {Leitherer}, C., {Garnett}, D.~R., \& {van der
  Rydt}, F. 1998, \apj, 503, 646, \dodoi{10.1086/306035}

\bibitem[{{Henry} {et~al.}(2018){Henry}, {Stephenson}, {Miller Bertolami},
  {Kwitter}, \& {Balick}}]{Henry+2018}
{Henry}, R.~B.~C., {Stephenson}, B.~G., {Miller Bertolami}, M.~M., {Kwitter},
  K.~B., \& {Balick}, B. 2018, \mnras, 473, 241, \dodoi{10.1093/mnras/stx2286}

\bibitem[{{Hern{\'a}ndez-P{\'e}rez} \& {Bruzual}(2014)}]{Hernandez+2014}
{Hern{\'a}ndez-P{\'e}rez}, F., \& {Bruzual}, G. 2014, \mnras, 444, 2571,
  \dodoi{10.1093/mnras/stu1627}

\bibitem[{{Herwig}(2000)}]{Herwig+2000}
{Herwig}, F. 2000, \aap, 360, 952

\bibitem[{{Hirschmann} {et~al.}(2017){Hirschmann}, {Charlot}, {Feltre}, {Naab},
  {Choi}, {Ostriker}, \& {Somerville}}]{Hirschmann+2017}
{Hirschmann}, M., {Charlot}, S., {Feltre}, A., {et~al.} 2017, \mnras, 472,
  2468, \dodoi{10.1093/mnras/stx2180}

\bibitem[{{Ho}(1999)}]{Ho+1999}
{Ho}, L. 1999, in Astrophysics and Space Science Library, Vol. 234,
  Observational Evidence for the Black Holes in the Universe, ed. S.~K.
  {Chakrabarti}, 157

\bibitem[{{Ho}(2009)}]{Ho+2009}
{Ho}, L.~C. 2009, \apj, 699, 638, \dodoi{10.1088/0004-637X/699/1/638}

\bibitem[{{James} \& {Percival}(2015)}]{James+2015}
{James}, P.~A., \& {Percival}, S.~M. 2015, \mnras, 450, 3503,
  \dodoi{10.1093/mnras/stv846}

\bibitem[{{Johansson} {et~al.}(2016){Johansson}, {Woods}, {Gilfanov}, {Sarzi},
  {Chen}, \& {Oh}}]{Johansson+2016}
{Johansson}, J., {Woods}, T.~E., {Gilfanov}, M., {et~al.} 2016, \mnras, 461,
  4505, \dodoi{10.1093/mnras/stw1668}

\bibitem[{Jones {et~al.}(2001--)Jones, Oliphant, Peterson, {et~al.}}]{SciPy}
Jones, E., Oliphant, T., Peterson, P., {et~al.} 2001--, {SciPy}: Open source
  scientific tools for {Python}.
\newblock \url{http://www.scipy.org/}

\bibitem[{{Kalirai} {et~al.}(2009){Kalirai}, {Saul Davis}, {Richer},
  {Bergeron}, {Catelan}, {Hansen}, \& {Rich}}]{Kalirai+2009}
{Kalirai}, J.~S., {Saul Davis}, D., {Richer}, H.~B., {et~al.} 2009, \apj, 705,
  408, \dodoi{10.1088/0004-637X/705/1/408}

\bibitem[{{Karakas}(2010)}]{Karakas+2010}
{Karakas}, A.~I. 2010, \mnras, 403, 1413,
  \dodoi{10.1111/j.1365-2966.2009.16198.x}

\bibitem[{{Kauffmann} {et~al.}(2003){Kauffmann}, {Heckman}, {Tremonti},
  {Brinchmann}, {Charlot}, {White}, {Ridgway}, {Brinkmann}, {Fukugita}, {Hall},
  {Ivezi{\'c}}, {Richards}, \& {Schneider}}]{Kauffmann+2003b}
{Kauffmann}, G., {Heckman}, T.~M., {Tremonti}, C., {et~al.} 2003, \mnras, 346,
  1055, \dodoi{10.1111/j.1365-2966.2003.07154.x}

\bibitem[{{Kaviraj} {et~al.}(2007){Kaviraj}, {Schawinski}, {Devriendt},
  {Ferreras}, {Khochfar}, {Yoon}, {Yi}, {Deharveng}, {Boselli}, {Barlow},
  {Conrow}, {Forster}, {Friedman}, {Martin}, {Morrissey}, {Neff},
  {Schiminovich}, {Seibert}, {Small}, {Wyder}, {Bianchi}, {Donas}, {Heckman},
  {Lee}, {Madore}, {Milliard}, {Rich}, \& {Szalay}}]{Kaviraj+2007}
{Kaviraj}, S., {Schawinski}, K., {Devriendt}, J.~E.~G., {et~al.} 2007, \apjs,
  173, 619, \dodoi{10.1086/516633}

\bibitem[{{Kehrig} {et~al.}(2012){Kehrig}, {Monreal-Ibero}, {Papaderos},
  {V{\'{\i}}lchez}, {Gomes}, {Masegosa}, {S{\'a}nchez}, {Lehnert}, {Cid
  Fernandes}, {Bland-Hawthorn}, {Bomans}, {Marquez}, {Mast}, {Aguerri},
  {L{\'o}pez-S{\'a}nchez}, {Marino}, {Pasquali}, {Perez}, {Roth},
  {S{\'a}nchez-Bl{\'a}zquez}, \& {Ziegler}}]{Kehrig+2012}
{Kehrig}, C., {Monreal-Ibero}, A., {Papaderos}, P., {et~al.} 2012, \aap, 540,
  A11, \dodoi{10.1051/0004-6361/201118357}

\bibitem[{{Kewley} {et~al.}(2001){Kewley}, {Dopita}, {Sutherland}, {Heisler},
  \& {Trevena}}]{Kewley+2001}
{Kewley}, L.~J., {Dopita}, M.~A., {Sutherland}, R.~S., {Heisler}, C.~A., \&
  {Trevena}, J. 2001, \apj, 556, 121, \dodoi{10.1086/321545}

\bibitem[{{Kewley} {et~al.}(2006){Kewley}, {Groves}, {Kauffmann}, \&
  {Heckman}}]{Kewley+2006}
{Kewley}, L.~J., {Groves}, B., {Kauffmann}, G., \& {Heckman}, T. 2006, \mnras,
  372, 961, \dodoi{10.1111/j.1365-2966.2006.10859.x}

\bibitem[{{Kormendy} \& {Ho}(2013)}]{Kormendy+2013}
{Kormendy}, J., \& {Ho}, L.~C. 2013, \araa, 51, 511,
  \dodoi{10.1146/annurev-astro-082708-101811}

\bibitem[{{Koski} \& {Osterbrock}(1976)}]{Koski+1976}
{Koski}, A.~T., \& {Osterbrock}, D.~E. 1976, \apjl, 203, L49,
  \dodoi{10.1086/182017}

\bibitem[{{Kroupa}(2001)}]{Kroupa+2001}
{Kroupa}, P. 2001, \mnras, 322, 231, \dodoi{10.1046/j.1365-8711.2001.04022.x}

\bibitem[{{Macchetto} {et~al.}(1996){Macchetto}, {Pastoriza}, {Caon}, {Sparks},
  {Giavalisco}, {Bender}, \& {Capaccioli}}]{Maccehetto+1996}
{Macchetto}, F., {Pastoriza}, M., {Caon}, N., {et~al.} 1996, \aaps, 120, 463

\bibitem[{{Maciel} {et~al.}(2017){Maciel}, {Costa}, \&
  {Cavichia}}]{Maciel+2017}
{Maciel}, W.~J., {Costa}, R.~D.~D., \& {Cavichia}, O. 2017, \rmxaa, 53, 151.
\newblock \doarXiv{1702.03721}

\bibitem[{{Maraston}(2005)}]{Maraston+2005}
{Maraston}, C. 2005, \mnras, 362, 799, \dodoi{10.1111/j.1365-2966.2005.09270.x}

\bibitem[{{Martin} {et~al.}(2005){Martin}, {Fanson}, {Schiminovich},
  {Morrissey}, {Friedman}, {Barlow}, {Conrow}, {Grange}, {Jelinsky},
  {Milliard}, {Siegmund}, {Bianchi}, {Byun}, {Donas}, {Forster}, {Heckman},
  {Lee}, {Madore}, {Malina}, {Neff}, {Rich}, {Small}, {Surber}, {Szalay},
  {Welsh}, \& {Wyder}}]{Martin+2005}
{Martin}, D.~C., {Fanson}, J., {Schiminovich}, D., {et~al.} 2005, \apjl, 619,
  L1, \dodoi{10.1086/426387}

\bibitem[{{McDonald} \& {Zijlstra}(2015)}]{McDonald+2015}
{McDonald}, I., \& {Zijlstra}, A.~A. 2015, \mnras, 448, 502,
  \dodoi{10.1093/mnras/stv007}

\bibitem[{{Miller Bertolami}(2016)}]{Miller+2016}
{Miller Bertolami}, M.~M. 2016, \aap, 588, A25,
  \dodoi{10.1051/0004-6361/201526577}

\bibitem[{{O'Connell}(1999)}]{OConnell+1999}
{O'Connell}, R.~W. 1999, \araa, 37, 603, \dodoi{10.1146/annurev.astro.37.1.603}

\bibitem[{{Oosterloo} {et~al.}(2010){Oosterloo}, {Morganti}, {Crocker},
  {J{\"u}tte}, {Cappellari}, {de Zeeuw}, {Krajnovi{\'c}}, {McDermid},
  {Kuntschner}, {Sarzi}, \& {Weijmans}}]{Oosterloo+2010}
{Oosterloo}, T., {Morganti}, R., {Crocker}, A., {et~al.} 2010, \mnras, 409,
  500, \dodoi{10.1111/j.1365-2966.2010.17351.x}

\bibitem[{{Pandya} {et~al.}(2017){Pandya}, {Greene}, {Ma}, {Veale}, {Ene},
  {Davis}, {Blakeslee}, {Goulding}, {McConnell}, {Nyland}, \&
  {Thomas}}]{Pandya+2017}
{Pandya}, V., {Greene}, J.~E., {Ma}, C.-P., {et~al.} 2017, \apj, 837, 40,
  \dodoi{10.3847/1538-4357/aa5ebc}

\bibitem[{{Papaderos} {et~al.}(2013){Papaderos}, {Gomes}, {V{\'{\i}}lchez},
  {Kehrig}, {Lehnert}, {Ziegler}, {S{\'a}nchez}, {Husemann}, {Monreal-Ibero},
  {Garc{\'{\i}}a-Benito}, {Bland-Hawthorn}, {Cortijo-Ferrero}, {de
  Lorenzo-C{\'a}ceres}, {del Olmo}, {Falc{\'o}n-Barroso}, {Galbany},
  {Iglesias-P{\'a}ramo}, {L{\'o}pez-S{\'a}nchez}, {Marquez}, {Moll{\'a}},
  {Mast}, {van de Ven}, \& {Wisotzki}}]{Papaderos+2013}
{Papaderos}, P., {Gomes}, J.~M., {V{\'{\i}}lchez}, J.~M., {et~al.} 2013, \aap,
  555, L1, \dodoi{10.1051/0004-6361/201321681}

\bibitem[{{Paxton} {et~al.}(2011){Paxton}, {Bildsten}, {Dotter}, {Herwig},
  {Lesaffre}, \& {Timmes}}]{Paxton+2011}
{Paxton}, B., {Bildsten}, L., {Dotter}, A., {et~al.} 2011, \apjs, 192, 3,
  \dodoi{10.1088/0067-0049/192/1/3}

\bibitem[{{Paxton} {et~al.}(2013){Paxton}, {Cantiello}, {Arras}, {Bildsten},
  {Brown}, {Dotter}, {Mankovich}, {Montgomery}, {Stello}, {Timmes}, \&
  {Townsend}}]{Paxton+2013}
{Paxton}, B., {Cantiello}, M., {Arras}, P., {et~al.} 2013, \apjs, 208, 4,
  \dodoi{10.1088/0067-0049/208/1/4}

\bibitem[{{Paxton} {et~al.}(2015){Paxton}, {Marchant}, {Schwab}, {Bauer},
  {Bildsten}, {Cantiello}, {Dessart}, {Farmer}, {Hu}, {Langer}, {Townsend},
  {Townsley}, \& {Timmes}}]{Paxton+2015}
{Paxton}, B., {Marchant}, P., {Schwab}, J., {et~al.} 2015, \apjs, 220, 15,
  \dodoi{10.1088/0067-0049/220/1/15}

\bibitem[{{Rauch}(2003)}]{Rauch+2003}
{Rauch}, T. 2003, \aap, 403, 709, \dodoi{10.1051/0004-6361:20030412}

\bibitem[{{Reimers}(1975)}]{Reimers+1975}
{Reimers}, D. 1975, Memoires of the Societe Royale des Sciences de Liege, 8,
  369

\bibitem[{{Rosenfield} {et~al.}(2012){Rosenfield}, {Johnson}, {Girardi},
  {Dalcanton}, {Bressan}, {Lang}, {Williams}, {Guhathakurta}, {Howley},
  {Lauer}, {Bell}, {Bianchi}, {Caldwell}, {Dolphin}, {Dorman}, {Gilbert},
  {Kalirai}, {Larsen}, {Olsen}, {Rix}, {Seth}, {Skillman}, \&
  {Weisz}}]{Rosenfield+2012}
{Rosenfield}, P., {Johnson}, L.~C., {Girardi}, L., {et~al.} 2012, \apj, 755,
  131, \dodoi{10.1088/0004-637X/755/2/131}

\bibitem[{{Rosenfield} {et~al.}(2014){Rosenfield}, {Marigo}, {Girardi},
  {Dalcanton}, {Bressan}, {Gullieuszik}, {Weisz}, {Williams}, {Dolphin}, \&
  {Aringer}}]{Rosenfield+2014}
{Rosenfield}, P., {Marigo}, P., {Girardi}, L., {et~al.} 2014, \apj, 790, 22,
  \dodoi{10.1088/0004-637X/790/1/22}

\bibitem[{{Rowlands} {et~al.}(2012){Rowlands}, {Dunne}, {Maddox}, {Bourne},
  {Gomez}, {Kaviraj}, {Bamford}, {Brough}, {Charlot}, {da Cunha}, {Driver},
  {Eales}, {Hopkins}, {Kelvin}, {Nichol}, {Sansom}, {Sharp}, {Smith}, {Temi},
  {van der Werf}, {Baes}, {Cava}, {Cooray}, {Croom}, {Dariush}, {de Zotti},
  {Dye}, {Fritz}, {Hopwood}, {Ibar}, {Ivison}, {Liske}, {Loveday}, {Madore},
  {Norberg}, {Popescu}, {Rigby}, {Robotham}, {Rodighiero}, {Seibert}, \&
  {Tuffs}}]{Rowlands+2012}
{Rowlands}, K., {Dunne}, L., {Maddox}, S., {et~al.} 2012, \mnras, 419, 2545,
  \dodoi{10.1111/j.1365-2966.2011.19905.x}

\bibitem[{{Sarzi} {et~al.}(2006){Sarzi}, {Falc{\'o}n-Barroso}, {Davies},
  {Bacon}, {Bureau}, {Cappellari}, {de Zeeuw}, {Emsellem}, {Fathi},
  {Krajnovi{\'c}}, {Kuntschner}, {McDermid}, \& {Peletier}}]{Sarzi+2006}
{Sarzi}, M., {Falc{\'o}n-Barroso}, J., {Davies}, R.~L., {et~al.} 2006, \mnras,
  366, 1151, \dodoi{10.1111/j.1365-2966.2005.09839.x}

\bibitem[{{Sarzi} {et~al.}(2010){Sarzi}, {Shields}, {Schawinski}, {Jeong},
  {Shapiro}, {Bacon}, {Bureau}, {Cappellari}, {Davies}, {de Zeeuw}, {Emsellem},
  {Falc{\'o}n-Barroso}, {Krajnovi{\'c}}, {Kuntschner}, {McDermid}, {Peletier},
  {van den Bosch}, {van de Ven}, \& {Yi}}]{Sarzi+2010}
{Sarzi}, M., {Shields}, J.~C., {Schawinski}, K., {et~al.} 2010, \mnras, 402,
  2187, \dodoi{10.1111/j.1365-2966.2009.16039.x}

\bibitem[{{Schawinski} {et~al.}(2007){Schawinski}, {Kaviraj}, {Khochfar},
  {Yoon}, {Yi}, {Deharveng}, {Boselli}, {Barlow}, {Conrow}, {Forster},
  {Friedman}, {Martin}, {Morrissey}, {Neff}, {Schiminovich}, {Seibert},
  {Small}, {Wyder}, {Bianchi}, {Donas}, {Heckman}, {Lee}, {Madore}, {Milliard},
  {Rich}, \& {Szalay}}]{Schawinski+2007}
{Schawinski}, K., {Kaviraj}, S., {Khochfar}, S., {et~al.} 2007, \apjs, 173,
  512, \dodoi{10.1086/516631}

\bibitem[{{Sharp} \& {Bland-Hawthorn}(2010)}]{Sharp+2010}
{Sharp}, R.~G., \& {Bland-Hawthorn}, J. 2010, \apj, 711, 818,
  \dodoi{10.1088/0004-637X/711/2/818}

\bibitem[{{Singh} {et~al.}(2013){Singh}, {van de Ven}, {Jahnke}, {Lyubenova},
  {Falc{\'o}n-Barroso}, {Alves}, {Cid Fernandes}, {Galbany},
  {Garc{\'{\i}}a-Benito}, {Husemann}, {Kennicutt}, {Marino}, {M{\'a}rquez},
  {Masegosa}, {Mast}, {Pasquali}, {S{\'a}nchez}, {Walcher}, {Wild}, {Wisotzki},
  \& {Ziegler}}]{Singh+2013}
{Singh}, R., {van de Ven}, G., {Jahnke}, K., {et~al.} 2013, \aap, 558, A43,
  \dodoi{10.1051/0004-6361/201322062}

\bibitem[{{Smith}(2014)}]{Smith+2014}
{Smith}, N. 2014, \araa, 52, 487, \dodoi{10.1146/annurev-astro-081913-040025}

\bibitem[{{Stasi{\'n}ska} {et~al.}(2008){Stasi{\'n}ska}, {Vale Asari}, {Cid
  Fernandes}, {Gomes}, {Schlickmann}, {Mateus}, {Schoenell}, {Sodr{\'e}}, \&
  {Seagal Collaboration}}]{Stasinska+2008}
{Stasi{\'n}ska}, G., {Vale Asari}, N., {Cid Fernandes}, R., {et~al.} 2008,
  \mnras, 391, L29, \dodoi{10.1111/j.1745-3933.2008.00550.x}

\bibitem[{{Taniguchi} {et~al.}(2000){Taniguchi}, {Shioya}, \&
  {Murayama}}]{Taniguchi+2000}
{Taniguchi}, Y., {Shioya}, Y., \& {Murayama}, T. 2000, \aj, 120, 1265,
  \dodoi{10.1086/301520}

\bibitem[{{Thomas} {et~al.}(2005){Thomas}, {Maraston}, {Bender}, \& {Mendes de
  Oliveira}}]{Thomas+2005}
{Thomas}, D., {Maraston}, C., {Bender}, R., \& {Mendes de Oliveira}, C. 2005,
  \apj, 621, 673, \dodoi{10.1086/426932}

\bibitem[{{Vassiliadis} \& {Wood}(1994)}]{Vassiliadis+1994}
{Vassiliadis}, E., \& {Wood}, P.~R. 1994, \apjs, 92, 125,
  \dodoi{10.1086/191962}

\bibitem[{{Weiss} \& {Ferguson}(2009)}]{Weiss+2009}
{Weiss}, A., \& {Ferguson}, J.~W. 2009, \aap, 508, 1343,
  \dodoi{10.1051/0004-6361/200912043}

\bibitem[{{Woods} \& {Gilfanov}(2014)}]{Woods+2014}
{Woods}, T.~E., \& {Gilfanov}, M. 2014, \mnras, 439, 2351,
  \dodoi{10.1093/mnras/stu072}

\bibitem[{{Yan}(2018{\natexlab{a}})}]{Yan+2018a}
{Yan}, R. 2018{\natexlab{a}}, \mnras, 481, 476, \dodoi{10.1093/mnras/sty2143}

\bibitem[{{Yan}(2018{\natexlab{b}})}]{Yan+2018b}
---. 2018{\natexlab{b}}, \mnras, 481, 467, \dodoi{10.1093/mnras/sty502}

\bibitem[{{Yan} \& {Blanton}(2012)}]{Yan+2012}
{Yan}, R., \& {Blanton}, M.~R. 2012, \apj, 747, 61,
  \dodoi{10.1088/0004-637X/747/1/61}

\bibitem[{{Yaron} {et~al.}(2017){Yaron}, {Prialnik}, {Kovetz}, \&
  {Shara}}]{Yaron+2017}
{Yaron}, O., {Prialnik}, D., {Kovetz}, A., \& {Shara}, M.~M. 2017, arXiv
  e-prints, arXiv:1709.02127.
\newblock \doarXiv{1709.02127}

\bibitem[{{Yi} {et~al.}(2005){Yi}, {Yoon}, {Kaviraj}, {Deharveng}, {Rich},
  {Salim}, {Boselli}, {Lee}, {Ree}, {Sohn}, {Rey}, {Lee}, {Rhee}, {Bianchi},
  {Byun}, {Donas}, {Friedman}, {Heckman}, {Jelinsky}, {Madore}, {Malina},
  {Martin}, {Milliard}, {Morrissey}, {Neff}, {Schiminovich}, {Siegmund},
  {Small}, {Szalay}, {Jee}, {Kim}, {Barlow}, {Forster}, {Welsh}, \&
  {Wyder}}]{Yi+2005}
{Yi}, S.~K., {Yoon}, S.-J., {Kaviraj}, S., {et~al.} 2005, \apjl, 619, L111,
  \dodoi{10.1086/422811}

\bibitem[{{Zhu} {et~al.}(2010){Zhu}, {Blanton}, \& {Moustakas}}]{Zhu+2010}
{Zhu}, G., {Blanton}, M.~R., \& {Moustakas}, J. 2010, \apj, 722, 491,
  \dodoi{10.1088/0004-637X/722/1/491}

\end{thebibliography}
\end{document}